\documentclass[reprint,aps,nofootinbib,superscriptaddress]{revtex4-2}
\usepackage{amsmath,latexsym,amssymb,amsfonts}
\usepackage{color,graphicx}
\usepackage{dcolumn}
\usepackage{bm}
\usepackage{subfloat}
\usepackage{caption}
\usepackage{subfig}

\def\red{\color{red}} 
\def\red{}          

\begin{document}

\title{\textbf{General relativistic effects on pulsar radiation}}

\author{Dong-Hoon Kim}
\email{ki13130@gmail.com} 
\affiliation{The Research Institute of Basic Sciences, Seoul National University, Seoul 08826,
Republic of Korea}
\affiliation{Department of Physics and Astronomy, Seoul National University, Seoul 08826,
Republic of Korea}
\author{Sascha Trippe}
\email{trippe@snu.ac.kr; corresponding author} 
\affiliation{Department of Physics and Astronomy, Seoul National University, Seoul 08826,
Republic of Korea}

\begin{abstract}
\centerline{\bf ABSTRACT} The magnetosphere of, and electromagnetic (EM)
radiation from pulsars are usually described in the framework of classical
electrodynamics. For some pulsars, however, whose emission heights are
relatively close to the surface of the neutron star, general relativistic
effects might modify the emission from the pulsar. We consider a magnetic
dipole model of a pulsar to investigate general relativistic effects on EM
radiation from it. Our study includes general relativistic modifications
applicable to some significant issues in pulsar astronomy, such as the
magnetosphere structure and pulse profiles. We implement computation of the
magnetic field in the pulsar magnetosphere from a solution to Maxwell's
equations defined in the strongly curved spacetime around a pulsar and find
that the field exhibits a strong gravitational effect. The effect modifies
curvature radiation of a pulsar, which then leads to modifications of the
pulse profiles of radio emission. We take the pulsar PSR J1828--1101 as an
example and work out Stokes parameters to simulate the pulse profiles for
its main and interpulse emissions theoretically, which exhibit the
gravitational effects clearly; however, their testability is beyond the
current detection capabilities, with the absolute magnitude of the pulse
profiles not being precisely predictable.
\end{abstract}

\maketitle

\section{Introduction\label{intro}}

Despite decades of study, pulsars still pose challenges for both
observations and theory. To first order, pulsars are described as
\textquotedblleft cosmic light houses\textquotedblright : neutron stars with
strong magnetic dipole fields rotating about an axis which is misaligned
with respect to the magnetic axis. Electrically charged particles are
accelerated along the magnetic field lines, especially around the magnetic
poles, and emit electromagnetic (EM) radiation across the entire EM spectrum
from radio to $\gamma $-rays. {\red If the rotation axis of the neutron star
and the magnetic axis are titled relative to each other, a distant observer
may observe characteristic radiation pulses.} Electrodynamics dictates that
the loss of energy due to radiation leads to a characteristic slowdown rate $%
\dot{P}$ of the pulsar period $P$, in good agreement with observations \cite%
{lorimer2008}. However, the complex physical processes involved in shaping
pulsar magnetospheres and radiation mechanisms -- particle acceleration,
non-thermal emission, pair creation, relativistic reconnection, and
(special) relativistic magnetohydrodynamics -- require exhaustive
theoretical and numerical studies to achieve a more complete understanding
of pulsar physics \cite{cerutti2017}. Understanding the emission from, and
brightness of, pulsars across the EM spectrum has consequences beyond
neutron star physics. One example for this is the observation of $\gamma $
radiation from the Galactic centre for which the annihilation of exotic dark
matter particles was suggested as source. Depending on the spectral
distribution of pulsar emission, radiation from a large population of
millisecond pulsars explains the signal equally well, thus eliminating the
need for exotic explanations \cite{yuan2014, eckner2018}.

One aspect that is usually neglected is the impact of general relativity on
the radio emission. Depending on where the emission takes place, gravity can
have a noticeable effect on the pulse profiles and the luminosity of a
pulsar from a theoretical perspective; therefore, its effect should be taken
into consideration for precise and accurate analyses of observational data.
Neutron stars have masses between 1.2 and 2.0 solar masses and radii of
about $11\,\mathrm{km}$ \cite{ozel2016}, corresponding to about two to three
Schwarzschild radii. This means that general relativistic effects are
important at least close to their surfaces. {\red The radiation received by
a distant observer is emitted at a specific ``emission height'' from the
centre of the neutron star: the peak frequency of curvature radiation,
assumed to be the observing frequency, is a function of particle Lorentz
factor and curvature radius; for a given frequency and Lorentz factor, the
curvature radius and thus the emission height follow \cite{Ganga2004}.} For
\textquotedblleft classical\textquotedblright\ pulsars with rotation periods
on the order of one second, the bulk of the EM radiation is emitted at
emission heights several hundred kilometers above the surface, corresponding
to roughly one hundred Schwarzschild radii \cite{ML2004}; this greatly
reduces relativistic effects and usually justifies neglecting them. Even
though, substantial relativistic effects may be expected at least for some
pulsars. A natural upper limit for the emission height is provided by the
light cylinder radius \cite{cerutti2017} which is 48~$\mathrm{km}$, about 10
Schwarzschild radii, for a pulsar with a period of one millisecond. This
suggests that millisecond pulsars are candidates for the detection of
relativistic effects in their radiation.\footnote{%
As of October 27, 2020, the number of pulsars listed in the ATFN Pulsar
Catalogue \cite{manchester2005} with $P\leq 5$\thinspace ms was 257, out of
a total of 2871 pulsars.} Emission heights vary substantially between
objects; radio observations have found emission heights below 100\thinspace $%
\mathrm{km}$ for several pulsars \cite{rankin2017, JK2019}, with the
smallest value being about 25\thinspace $\mathrm{km}$ for the millisecond
pulsar J1022+1001 \cite{rankin2017}.

Previous studies of general relativistic effects were mostly focused on the
EM field geometry around a rotating neutron star. Sengupta \cite%
{Sengupta1995} investigated the importance of general relativistic
corrections to the induced electric field exterior to a neutron star by
considering the simplest aligned vacuum and nonvacuum magnetosphere models.
Konno and Kojima \cite{KK2000} worked out EM fields around a rotating star
endowed with an aligned dipole magnetic field by solving Maxwell's equations
in a slowly rotating Kerr spacetime. Rezzolla et al. \cite{RAM2001} obtained
analytic solutions of Maxwell's equations in the internal and external
background spacetime of a slowly rotating misaligned magnetized neutron
star. Ruiz et al. \cite{Ruiz2014} used numerical simulations to estimate the
general relativistic spin-down luminosity of pulsars. P\'{e}tri \cite%
{Petri2016} performed time-dependent simulations of Maxwell's equations in a
stationary background metric in the slow-rotation approximation and found
that the Poynting flux observed at a large distance is substantially higher
in the general relativistic case compared to the non-relativistic case.

In this work, we investigate general relativistic modifications of pulsar
radiation analytically, by considering a magnetic dipole model which is an
`oblique' (or misaligned) rotator with an inclination angle. We derive
expressions for a general relativistic description of the pulsar
magnetosphere and the corresponding pulsar energetics and EM radiation.
Largely, our analysis proceeds in two steps through Sections \ref{EDP} and %
\ref{emission}. In Section \ref{EDP}, general relativistic modifications of
the magnetic field around a\ rotating neutron star are studied. To this end,
we specially prescribe a rotating Schwarzschild geometry, which is modified
from a slowly rotating Kerr spacetime but well suited to solving Maxwell's
equations for the magnetic field exterior to an oblique rotator. The results
obtained in Section \ref{EDP} are fully employed to develop our analysis for
EM radiation in Section \ref{emission}, such as very low-frequency magnetic
dipole radiation for pulsar spin-down and curvature radiation from the
pulsar PSR J1828--1101.

\section{Electrodynamics in the Pulsar Magnetosphere\label{EDP}}

\subsection{Geometric Configuration of an Obliquely Rotating Pulsar\label%
{config}}

We consider a magnetic dipole model of a pulsar in a general configuration:
a rotating neutron star having the magnetic field around it, which can be
well-approximated to be that of a pure magnetic dipole \cite%
{Sengupta1995,Petterson1974}, with its magnetic axis tilted from its
rotation axis by an angle $\alpha $. The spacetime geometry around the
pulsar is strongly curved due to the immense gravitation of the neutron
star. An obliquely rotating pulsar can be modeled using the Kerr geometry 
\cite{RAM2001}; more effectively, the geometry can be described by a Kerr
metric in the slow rotation limit:%
\begin{eqnarray}
ds^{2} &=&-f\left( r\right) c^{2}dt^{2}+\frac{1}{f\left( r\right) }%
dr^{2}+r^{2}d\theta ^{2}+r^{2}\sin ^{2}\theta d\tilde{\phi}^{2}  \notag \\
&&+\mathcal{O}\left( a^{2}\right) ,  \label{g1}
\end{eqnarray}%
where $f\left( r\right) \equiv 1-2GM/\left( c^{2}r\right) $, with $G$ being
the gravitational constant, $c$ being the speed of light and $M$ being the
total mass of the neutron star, and 
\begin{equation}
\tilde{\phi}\equiv \phi -\Omega _{\mathrm{f.d.}}t,  \label{cr1}
\end{equation}%
where 
\begin{equation}
\Omega _{\mathrm{f.d.}}=2GJ/\left( c^{2}r^{3}\right) =2GMa/\left(
cr^{3}\right)  \label{fd}
\end{equation}%
is the \textquotedblleft frame-dragging\textquotedblright\ angular frequency
of the pulsar with $J$ being the angular momentum and $a=J/\left( Mc\right) $
being the Kerr parameter. Here $\mathcal{O}\left( a^{2}\right) \sim \mathcal{%
O}\left( \Omega _{\mathrm{f.d.}}^{2}\right) $ represents any higher order
terms than linear in $a\sim \Omega _{\mathrm{f.d.}}$, which would come from
the full Kerr metric when expanded with respect to $a$, can be cast off in
the slow rotation limit.

\begin{figure}[tbp]
\centering
\includegraphics[width=8.6cm]{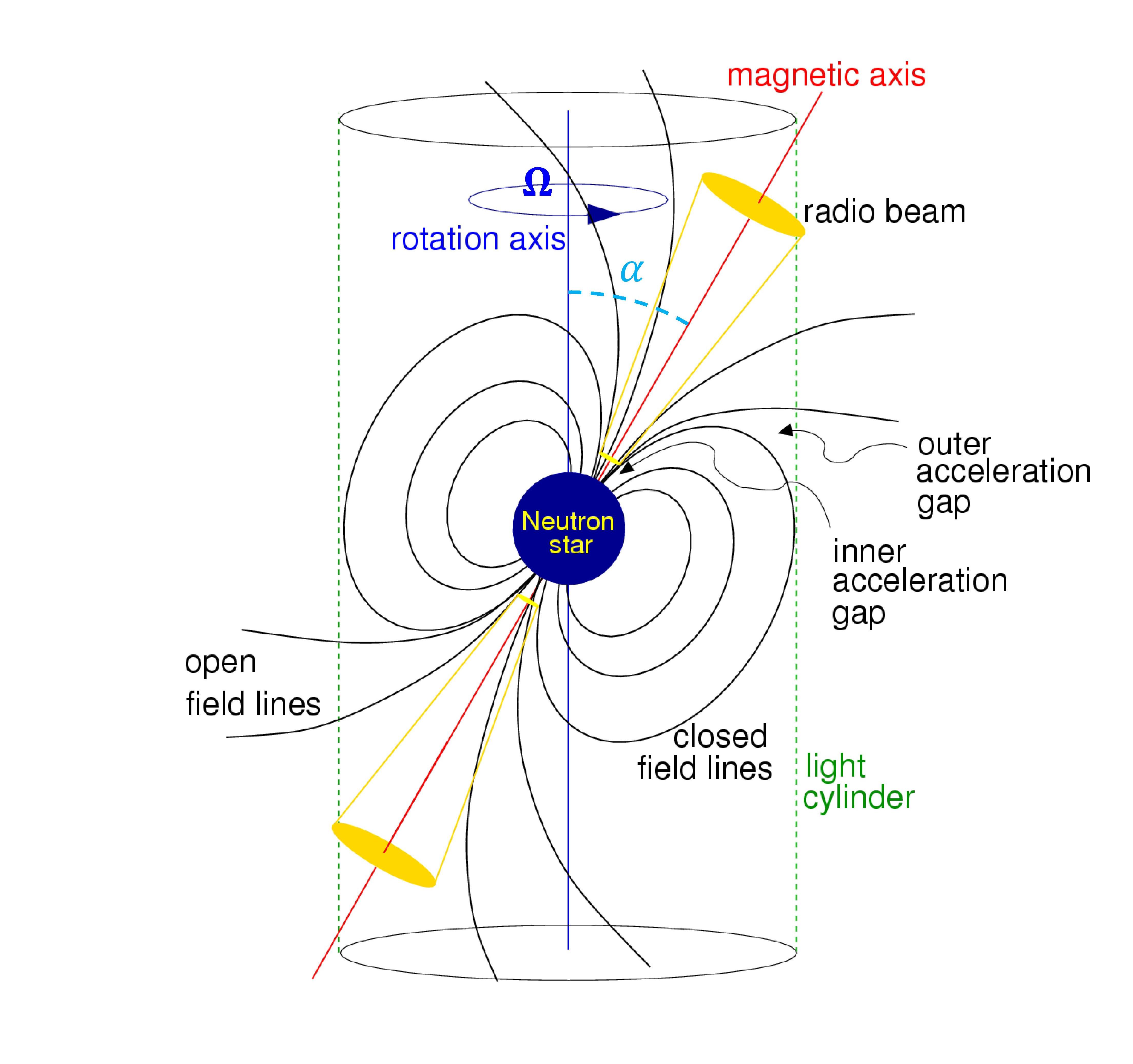}
\caption{A graphical illustration of our pulsar model: the magnetic axis is
inclined from the rotation axis by the angle $\protect\alpha $. (Credit:
Lorimer and Kramer \protect\cite{LK2005}, reproduced with some
modifications.) }
\label{fig1}
\end{figure}

Our pulsar model is graphically illustrated in Fig. \ref{fig1}. Note that
the magnetic axis, which coincides with the magnetic dipole and the
radiation beam direction, is inclined from the rotation axis by the angle $%
\alpha $. Also, for the rest of our analysis, it should be noted that $%
\Omega =\left\vert \mathbf{\Omega }\right\vert =2\pi /P$ denotes the pulsar
rotation frequency, which should be distinguished from the frame-dragging\
frequency $\Omega _{\mathrm{f.d.}}$ as defined by Eq. (\ref{fd}) above.
While $\Omega _{\mathrm{f.d.}}$ drops rapidly as one moves away from the
pulsar surface, $\Omega $ is a \textit{nearly} \textit{fixed} value through
the measurement of the rotation period $P$. One may express the angular
momentum $J$ as 
\begin{equation}
J=k^{2}MR_{\ast }^{2}\Omega ,  \label{J}
\end{equation}%
where $k^{2}\simeq 0.4$ depending on the neutron star equation of state and $%
R_{\ast }$ denotes the radius of the star \cite{Beskin2003}. Using this, we
can relate the frame-dragging\ frequency $\Omega _{\mathrm{f.d.}}$ to the
rotation frequency $\Omega $ \cite{Beskin2003}: 
\begin{equation}
\Omega _{\mathrm{f.d.}}=\Omega _{\ast }\left( R_{\ast }/r\right) ^{3},
\label{fd1}
\end{equation}%
where 
\begin{equation}
\Omega _{\ast }\equiv \Omega _{\mathrm{f.d.}}\left( r=R_{\ast }\right) =%
\left[ 2k^{2}GM/\left( c^{2}R_{\ast }\right) \right] \Omega  \label{om}
\end{equation}%
is the frame-dragging\ frequency defined at the stellar surface given by
means of Eqs. (\ref{fd}) and (\ref{J}).\footnote{%
A neutron star has a hard surface while a black hole has an event horizon.
With this distinction, a rotating neutron star is well-characterized by the
frame-dragging angular frequency evaluated at its surface.}

\subsection{EM Fields of an Obliquely Rotating Pulsar\label{oblique}}

In Section \ref{config} we have described the configuration of our pulsar:
the main features are (i) the star's magnetic axis is tilted from its
rotation axis by the angle $\alpha $, (ii) the star is slowly rotating at
the frequency $\Omega =2\pi /P$, where $P$ is the rotation period obtained
through measurement. Now, the spacetime geometry described by Eq. (\ref{g1})
is corotating slowly at the frame-dragging frequency $\Omega _{\mathrm{f.d.}%
} $, which varies with the radial position $r$ as given by Eq. (\ref{fd}),
rather than the pulsar rotation frequency $\Omega $. Analytical solutions of
Maxwell's equations in this geometry are available in the literature; e.g., 
\cite{RAM2001}. However, in the present analysis, we are concerned only with
the magnetic field outside the neutron star, which corresponds to part of
the stationary exterior vacuum solution of Maxwell's equations for a
misaligned magnetized rotator as obtained by \cite{RAM2001}. From the
solution, it should be noted that general relativistic corrections due to
the frame-dragging effect are not present in the expression for the magnetic
field in the slow rotation limit; that is, the expression is given in terms
of the pulsar rotation frequency $\Omega $ but not the frame-dragging
frequency $\Omega _{\mathrm{f.d.}}$. Then one might consider that the same
result can be obtained by solving Maxwell's equations for the magnetic field
in a frame that corotates with the pulsar at the frequency $\Omega $ and
then performing the coordinate transformation between this frame and the
inertial frame. This alternative method, albeit much simpler to implement,
is fully legitimate as long as our attention is restricted to the magnetic
field exterior to the slowly rotating star. Also, it is consistent with our
analysis for curvature radiation developed in Section \ref{radio}.\footnote{%
In order to describe curvature radiation, we introduce quantities such as
the curvature radius of a charge's trajectory along a magnetic field line,
the Stokes parameters defined from the Li\'{e}nard--Wiechert EM field due to
a charge moving along a magnetic field line, etc., which are derived in the
corotating frame first and then transformed to the inertial frame.} The
method is described in detail below.

\subsubsection{Transformation between the Inertial Frame and the Corotating
Frame\label{trans}}

Following the argument regarding the magnetic field exterior to the slowly
rotating star as given above, we can simplify our analysis by replacing $%
\Omega _{\mathrm{f.d.}}$ with the pulsar rotation frequency $\Omega =2\pi /P$%
, which is in correspondence with observation through the measurement of the
rotation period $P$. As $\tilde{\phi}$ is modified from Eq. (\ref{cr1}): 
\begin{equation}
\tilde{\phi}=\phi -\Omega t,  \label{cr2}
\end{equation}%
where $\Omega _{\mathrm{f.d.}}$ has been replaced by $\Omega $, the new
geometry modified from Eq. (\ref{g1}) is spherically symmetric apart from $%
\mathcal{O}\left( a^{2}\right) \sim \mathcal{O}\left( R_{\ast }\left(
v_{\ast }/c\right) ^{2}\right) $, where $v_{\ast }\equiv \Omega R_{\ast }$
is the star's spinning velocity at its surface.

\begin{figure}[tbp]
\centering
\includegraphics[width=8.6cm]{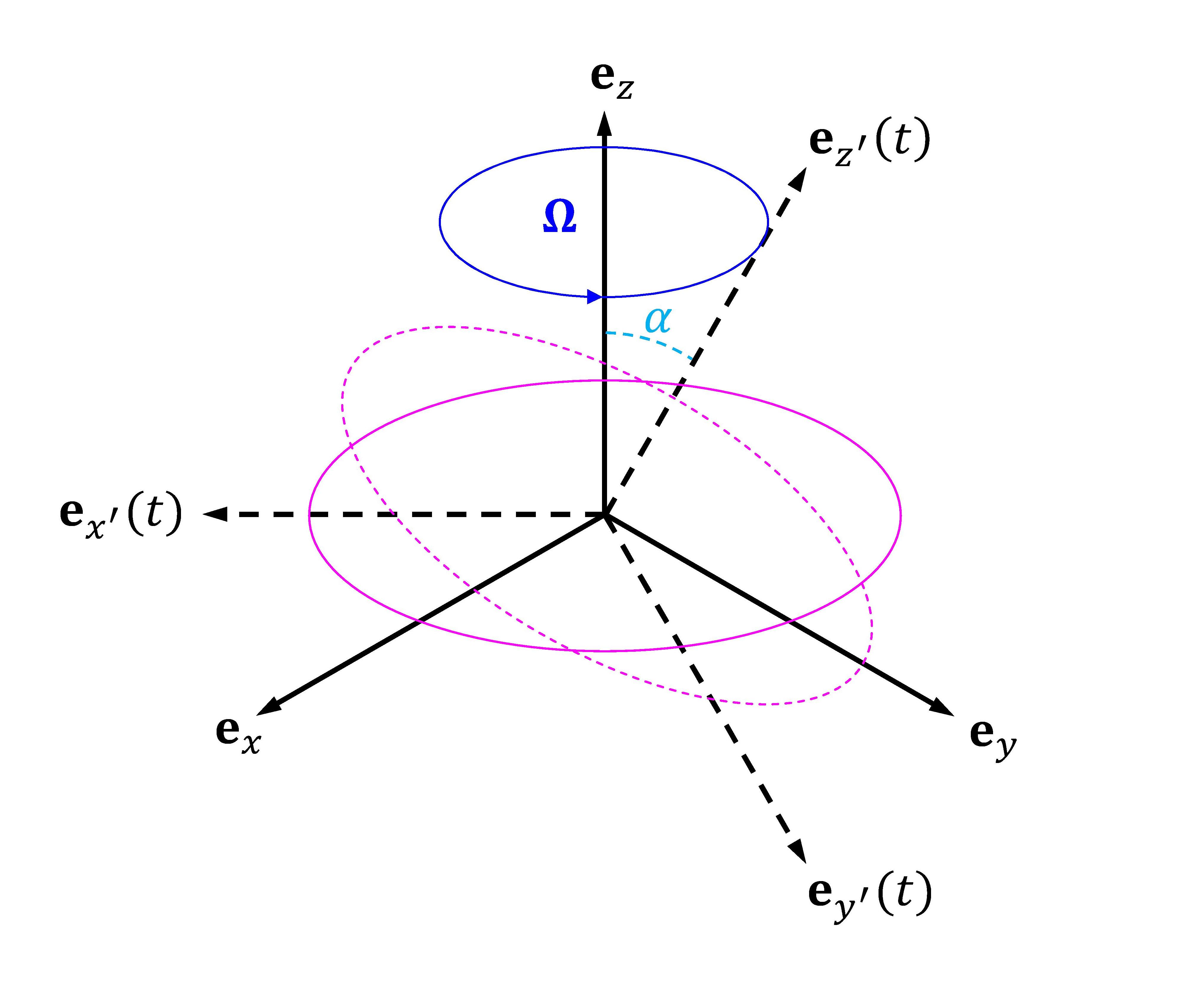}
\caption{The inertial frame $\mathbf{x}\equiv \left( x,y,z\right) $ and the
obliquely corotating frame $\mathbf{x}^{\prime }\equiv \left( x^{\prime
},y^{\prime },z^{\prime }\right) $. The $z^{\prime }$-axis of the latter
rotates around the $z$-axis of the former at the angular frequency $\Omega $%
, keeping the angle $\protect\alpha $ between the two axes. The relation
between the two frames is described mathematically by Eqs. (\protect\ref{x1}%
)-(\protect\ref{z1}) in Appendix \protect\ref{appA}.}
\label{fig2}
\end{figure}

The spherical symmetry of the modified geometry is maintained under the
transformation of coordinates from a frame $\left( t,r,\theta ,\tilde{\phi}%
\right) $ to another $\left( t,r,\theta ^{\prime },\phi ^{\prime }\right) $
on the 2-sphere. Suppose that $\left( x,y,z\right) $ and $\left( x^{\prime
},y^{\prime },z^{\prime }\right) $ are the Cartesian frames converted from
the spherical frames $\left( r,\theta ,\phi \right) $ and $\left( r,\theta
^{\prime },\phi ^{\prime }\right) $, respectively: following the usual
convention, $\theta $[$\theta ^{\prime }$] is defined as the polar angle
measured from the $z$[$z^{\prime }$]-axis while $\phi $[$\phi ^{\prime }$]
is the azimuthal angle around the $z$[$z^{\prime }$]-axis, with the $%
z^{\prime }$-axis being tilted from the $z$-axis by the angle $\alpha $. As
illustrated in Fig. \ref{fig1}, we may set the rotation axis and the
magnetic axis of our pulsar to be along the $z$-axis and the $z^{\prime }$%
-axis, respectively. Then it follows that the frame $\left( x^{\prime
},y^{\prime },z^{\prime }\right) $ is rotated relative to the frame $\left(
x,y,z\right) $ about the $z$-axis by the amount of rotation $\Omega t$,
while keeping the angle $\alpha $ between the $z$-axis and the $z^{\prime }$%
-axis. Fig. \ref{fig2} illustrates how these two frames are related to each
other. The transformation between the frames $\left( r,\theta ,\phi \right) $
and $\left( r,\theta ^{\prime },\phi ^{\prime }\right) $ can be determined
by way of the transformation between the frames $\left( x,y,z\right) $ and $%
\left( x^{\prime },y^{\prime },z^{\prime }\right) $. This would require a
somewhat involved analysis, and therefore we present only the result here
and the technical details are provided in Appendix \ref{appA}. With the
basis vectors $\left( \mathbf{e}_{\hat{r}},\mathbf{e}_{\hat{\theta}},\mathbf{%
e}_{\hat{\phi}}\right) $ and $\left( \mathbf{e}_{\hat{r}},\mathbf{e}_{\hat{%
\theta}^{\prime }},\mathbf{e}_{\hat{\phi}^{\prime }}\right) $ for the frames 
$\left( r,\theta ,\phi \right) $ and $\left( r,\theta ^{\prime },\phi
^{\prime }\right) $, respectively, it is shown that the two frames are
related to each other: 
\begin{equation}
\left[ 
\begin{array}{c}
\mathbf{e}_{\hat{r}} \\ 
\mathbf{e}_{\hat{\theta}} \\ 
\mathbf{e}_{\hat{\phi}}%
\end{array}%
\right] =\mathbf{M}\left[ 
\begin{array}{c}
\mathbf{e}_{\hat{r}} \\ 
\mathbf{e}_{\hat{\theta}^{\prime }} \\ 
\mathbf{e}_{\hat{\phi}^{\prime }}%
\end{array}%
\right] ,  \label{bv}
\end{equation}%
where%
\begin{widetext}
\begin{equation}
\mathbf{M}=\left[ 
\begin{array}{ccc}
1 & 0 & 0 \\ 
0 & \frac{\cos \alpha \sin \theta -\sin \alpha \cos \theta \cos \left( \phi
-\Omega t\right) }{\sin \theta ^{\prime }} & \frac{-\sin \alpha \sin \left(
\phi -\Omega t\right) }{\sin \theta ^{\prime }} \\ 
0 & \frac{\sin \alpha \sin \left( \phi -\Omega t\right) }{\sin \theta
^{\prime }} & \frac{\cos \alpha \sin \theta -\sin \alpha \cos \theta \cos
\left( \phi -\Omega t\right) }{\sin \theta ^{\prime }}%
\end{array}%
\right] ,  \label{M}
\end{equation}
\end{widetext}with $\theta ^{\prime }$ being expressed via 
\begin{equation}
\cos \theta ^{\prime }=\cos \alpha \cos \theta +\sin \alpha \sin \theta \cos
\left( \phi -\Omega t\right) .  \label{col}
\end{equation}%
This gives a definition of the \textquotedblleft magnetic
colatitude\textquotedblright\ \cite{MY2012}; that is, the corotating polar
angle evaluated in the inertial frame.

Letting $\left( \mathbf{e}_{\hat{r}},\mathbf{e}_{\hat{\theta}},\mathbf{e}_{%
\hat{\phi}}\right) $ and $\left( \mathbf{e}_{\hat{r}},\mathbf{e}_{\hat{\theta%
}^{\prime }},\mathbf{e}_{\hat{\phi}^{\prime }}\right) $ refer to the
inertial frame and the corotating frame, respectively, Eq. (\ref{M})
represents the transformation between these frames. Later in Section \ref%
{reference}, this will be employed to determine the EM fields for our
obliquely rotating pulsar as viewed by an observer in the inertial frame.

\subsubsection{Solving Maxwell's Equations in the Corotating Frame\label%
{Maxwell}}

We may identify the magnetic axis of our pulsar with the $z^{\prime }$-axis
as described in Section \ref{trans}. Now, taking advantage of the spherical
symmetry, the new geometry modified from (\ref{g1}), with (\ref{cr1}) being
replaced by (\ref{cr2}) can be reformed to another Schwarzschild metric
apart from $\mathcal{O}\left( a^{2}\right) $: 
\begin{eqnarray}
ds^{2} &=&-f\left( r\right) c^{2}dt^{2}+\frac{1}{f\left( r\right) }%
dr^{2}+r^{2}d\theta ^{\prime 2}+r^{2}\sin ^{2}\theta ^{\prime }d\phi
^{\prime 2}  \notag \\
&&+\mathcal{O}\left( a^{2}\right) ,  \label{g2}
\end{eqnarray}%
where $\theta ^{\prime }$ is the polar angle measured from the $z^{\prime }$%
-axis, i.e., magnetic axis and $\phi ^{\prime }$ is the azimuthal angle
around this axis. For an observer sitting in the frame $\left( t,r,\theta
^{\prime },\phi ^{\prime }\right) $, electromagnetism in our pulsar
magnetosphere will be described by the Maxwell's equations in the geometry (%
\ref{g2}):%
\begin{equation}
\nabla ^{2}A_{a}-R_{a}{}^{b}A_{b}=-\frac{4\pi }{c}j_{a},  \label{M1}
\end{equation}%
where $\nabla ^{2}$ denotes the curved spacetime d'Alembertian,$\ A_{a}$
represents the EM vector potential,$\ j_{a}$ is the current density, $%
R_{ab}{}$ is the Ricci tensor, and the indices $a$, $b$, ... refer to the
coordinates $\left( t,r,\theta ^{\prime },\phi ^{\prime }\right) $.

Following Petterson \cite{Petterson1974}, we can express the current density
in Eq. (\ref{M1}) as 
\begin{equation}
j^{\phi ^{\prime }}=\left( 1-\frac{2m}{r}\right) ^{1/2}\frac{\delta \left(
r-R_{\mathrm{o}}\right) \delta \left( \cos \theta ^{\prime }\right) I}{r^{2}}%
,  \label{j1}
\end{equation}%
where $m\equiv GM/c^{2}$ and $I$ is a static current loop of radius $R_{%
\mathrm{o}}$, encircling the equator of the neutron star, i.e., $\theta
^{\prime }=\pi /2$.\footnote{$I$ is the total current through the $\hat{r}%
\hat{\theta}^{\prime }$-plane as defined in the local Lorentz frame: 
\begin{equation*}
I=\iint j^{\hat{\phi}^{\prime }}d\hat{r}d\hat{\theta}^{\prime },
\end{equation*}%
where $j^{\hat{\phi}^{\prime }}=j^{\phi ^{\prime }}\left( \lambda _{\hat{\phi%
}^{\prime }}^{\phi ^{\prime }}\right) ^{-1}$, $d\hat{r}=dr\left( \lambda _{%
\hat{r}}^{r}\right) ^{-1}$ and $d\hat{\theta}^{\prime }=d\theta \left(
\lambda _{\hat{\theta}^{\prime }}^{\theta ^{\prime }}\right) ^{-1}$ by means
of the tetrad (\ref{te}) \cite{Petterson1974}.} Then a solution to Eq. (\ref%
{M1}) is obtained as \cite{Petterson1974,Ginzburg1964,Anderson1970} 
\begin{equation}
A_{\phi ^{\prime }}=-\frac{3\mu \sin ^{2}\theta ^{\prime }}{8m^{3}}\left[
r^{2}\ln \left( 1-\frac{2m}{r}\right) +2mr\left( 1+\frac{m}{r}\right) \right]
,  \label{A1}
\end{equation}%
where $\mu $ is the magnetic dipole moment associated with the current
density $j_{\phi ^{\prime }}$, and leads to \cite{Petterson1974}\footnote{%
A general relativistic effect is taken into consideration in this
expression: it depends on the mass of the star $M$ through $m=GM/c^{2}$. For
example, for a standard neutron star with $R_{\ast }\simeq 4GM/c^{2}$, we
have $\mu \simeq \pi R_{\ast }^{2}I/\sqrt{2}$ with $R_{\mathrm{o}}=R_{\ast }$
substituted into the expression, which contrasts with $\mu =\pi R_{\ast
}^{2}I$, the value in the flat spacetime limit $m\rightarrow 0$. On the
other hand, we find that $\mu $ vanishes in the limit $R_{\ast }\rightarrow
2GM/c^{2}$, i.e., approaching the event horizon. This corresponds to the
vanishing of the magnetic field for an observer at infinity when the source
approaches the horizon. This notion is in accordance with a theorem by Price 
\cite{Price1972}: during the process of gravitational collapse, all EM
multipole moments of the collapsing matter, except the electric monopole
moment, must disappear \cite{Petterson1974,Price1972}.} 
\begin{equation}
\mu =\pi R_{\mathrm{o}}^{2}\left( 1-\frac{2m}{R_{\mathrm{o}}}\right) ^{1/2}I.
\label{mu}
\end{equation}

In the same frame, using Eq. (\ref{A1}), we obtain the components of the
magnetic field from the definition of the EM field-strength tensor, $%
F_{ab}\equiv A_{b,a}-A_{a,b}$, where the comma followed by the subscript
indicates partial differentiation with respect to the subscript \cite%
{Sengupta1995,Petterson1974}:%
\begin{align}
& B_{\hat{r}}=F_{\hat{\theta}^{\prime }\hat{\phi}^{\prime }}=F_{\theta
^{\prime }\phi ^{\prime }}\lambda _{\hat{\theta}^{\prime }}^{\theta ^{\prime
}}\lambda _{\hat{\phi}^{\prime }}^{\phi ^{\prime }}  \notag \\
& =-\frac{3\mu \cos \theta ^{\prime }}{4m^{3}}\left[ \ln \left( 1-\frac{2m}{r%
}\right) +\frac{2m}{r}\left( 1+\frac{m}{r}\right) \right] ,  \label{B1} \\
& B_{\hat{\theta}^{\prime }}=F_{\hat{r}\hat{\phi}^{\prime }}=F_{r\phi
^{\prime }}\lambda _{\hat{r}}^{r}\lambda _{\hat{\phi}^{\prime }}^{\phi
^{\prime }}  \notag \\
& =\frac{3\mu \sin \theta ^{\prime }}{4m^{3}}\left[ \ln \left( 1-\frac{2m}{r}%
\right) +\frac{m}{r}\left( \left( 1-\frac{2m}{r}\right) ^{-1}+1\right) %
\right]  \notag \\
& \times \left( 1-\frac{2m}{r}\right) ^{1/2},  \label{B2}
\end{align}%
where the indices with a `hat' ( $\widehat{}$ ) denote the components in a
local Lorentz frame and$\ \lambda _{\hat{b}}^{a}$ represents the orthonormal
tetrad of the local Lorentz frame for the geometry (\ref{g2}), the
non-vanishing components of which are given by 
\begin{align}
& \left\{ \lambda _{\hat{t}}^{t},\lambda _{\hat{r}}^{r},\lambda _{\hat{\theta%
}^{\prime }}^{\theta ^{\prime }},\lambda _{\hat{\phi}^{\prime }}^{\phi
^{\prime }}\right\}  \notag \\
& =\left\{ \left( 1-\frac{2m}{r}\right) ^{-1/2},\left( 1-\frac{2m}{r}\right)
^{1/2},\frac{1}{r},\frac{1}{r\sin \theta ^{\prime }}\right\} .  \label{te}
\end{align}%
Dropping the $^{\prime }$ signs from this, the orthonormal tetrad defined in
the inertial frame $\left( t,r,\theta ,\phi \right) $ is also expressed in
the same manner.

It should be noted that the magnetic field components given by Eqs. (\ref{B1}%
) and (\ref{B2}) above would be the values as observed in the corotating
frame, and hence that they do not exhibit any effect of rotation of our
star; as if the star were static, with its spacetime geometry being
described by Eq. (\ref{g2}). Next, we will find how the star's rotation
affects the EM fields observed in the inertial frame.

\subsubsection{EM Fields as Observed in the Inertial Frame \label{reference}}

For an observer in the inertial frame, electromagnetism of the pulsar should
be different from that observed in the corotating frame as the rotation of
the star affects our observation of the EM fields.

First, the vector potential (\ref{A1}) should be transformed from the
corotating frame to the inertial frame: 
\begin{align}
& A_{\theta }=A_{\phi ^{\prime }}\lambda _{\hat{\phi}^{\prime }}^{\phi
^{\prime }}M_{\hat{\theta}}^{\hat{\phi}^{\prime }}\left( \lambda _{\hat{%
\theta}}^{\theta }\right) ^{-1}  \notag \\
& =\frac{3\mu \sin \alpha \sin \left( \phi -\Omega t\right) }{8m^{3}}  \notag
\\
& \times \left[ r^{2}\ln \left( 1-\frac{2m}{r}\right) +2mr\left( 1+\frac{m}{r%
}\right) \right] ,  \label{A2} \\
& A_{\phi }=A_{\phi ^{\prime }}\lambda _{\hat{\phi}^{\prime }}^{\phi
^{\prime }}M_{\hat{\phi}}^{\hat{\phi}^{\prime }}\left( \lambda _{\hat{\phi}%
}^{\phi }\right) ^{-1}  \notag \\
& =-\frac{3\mu \sin \theta \left[ \cos \alpha \sin \theta -\sin \alpha \cos
\theta \cos \left( \phi -\Omega t\right) \right] }{8m^{3}}  \notag \\
& \times \left[ r^{2}\ln \left( 1-\frac{2m}{r}\right) +2mr\left( 1+\frac{m}{r%
}\right) \right] ,  \label{A3}
\end{align}%
where $M_{\hat{a}}^{\hat{b}}$ is to be read off from the transformation
matrix given by Eq. (\ref{M}), and $\lambda _{\hat{b}}^{a}$ is given by Eq. (%
\ref{te}).

\begin{figure*}[tbp]
\centering
\includegraphics[width=16.5cm]{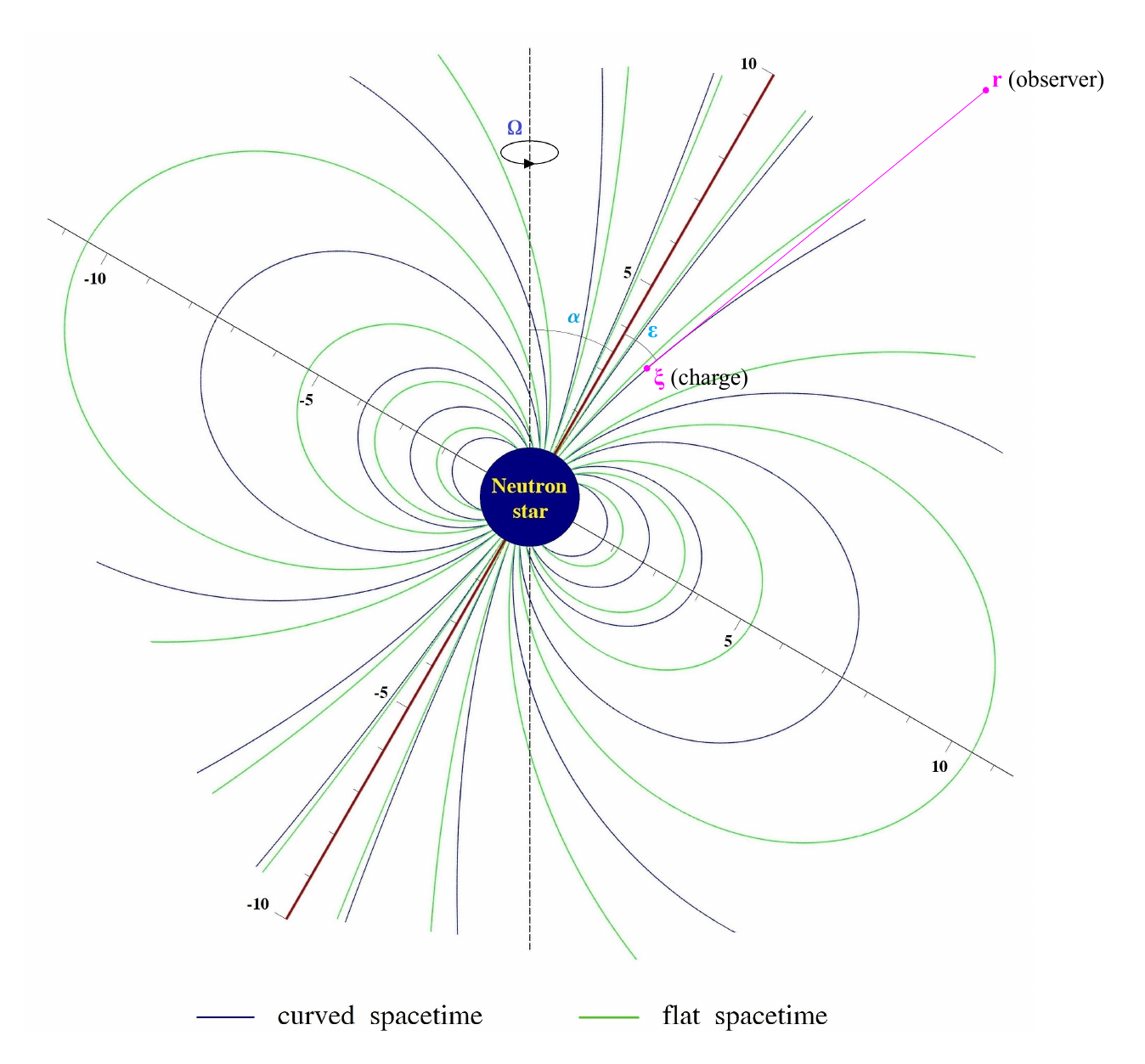}
\caption{The dipole magnetic field lines around the neutron star based on
Eqs. (\protect\ref{B3})-(\protect\ref{B5}), with $\protect\phi -\Omega t $
set to be $0$. The vertical dashed line (black) and the inclined solid line
(red) represent the rotation axis and the magnetic axis, respectively. The
field lines in curved spacetime (blue) are distinguished from those in flat
spacetime (green). The scale of the unity in this graph is equivalent to the
radius of the neutron star, $R_{\ast }\simeq 10^{6}\,\mathrm{cm}$. The
magenta line connecting a source charge ($\mathbf{\protect\xi }$) and the
observer ($\mathbf{r}$) illustrates a line of sight. While $\protect\alpha $
(between the rotation axis and the magnetic axis) denotes the inclination
angle, $\protect\varepsilon $ (between the magnetic axis and the line of
sight) denotes the sight line impact angle. }
\label{fig3}
\end{figure*}

The magnetic field in the inertial frame can be determined through $%
F_{ab}=A_{b,a}-A_{a,b}$, together with Eqs. (\ref{A2}) and (\ref{A3}) and
using Eq. (\ref{te}):\footnote{%
The same results are obtained by directly transforming the magnetic field
from the corotating frame to the reference frame by means of (\ref{B1}), (%
\ref{B2}) and (\ref{e4}).} 
\begin{align}
& B_{\hat{r}}=F_{\hat{\theta}\hat{\phi}}=\left( A_{\phi ,\theta }-A_{\theta
,\phi }\right) \lambda _{\hat{\theta}}^{\theta }\lambda _{\hat{\phi}}^{\phi }
\notag \\
& =-\frac{3\mu \left[ \cos \alpha \cos \theta +\sin \alpha \sin \theta \cos
\left( \phi -\Omega t\right) \right] }{4m^{3}}  \notag \\
& \times \left[ \ln \left( 1-\frac{2m}{r}\right) +\frac{2m}{r}\left( 1+\frac{%
m}{r}\right) \right] ,  \label{B3} \\
& B_{\hat{\theta}}=F_{\hat{\phi}\hat{r}}=-A_{\phi ,r}\lambda _{\hat{\phi}%
}^{\phi }\lambda _{\hat{r}}^{r}  \notag \\
& =\frac{3\mu \left[ \cos \alpha \sin \theta -\sin \alpha \cos \theta \cos
\left( \phi -\Omega t\right) \right] }{4m^{3}}  \notag \\
& \times \left[ \ln \left( 1-\frac{2m}{r}\right) +\frac{m}{r}\left( \left( 1-%
\frac{2m}{r}\right) ^{-1}+1\right) \right]  \notag \\
& \times \left( 1-\frac{2m}{r}\right) ^{1/2},  \label{B4} \\
& B_{\hat{\phi}}=F_{\hat{r}\hat{\theta}}=A_{\theta ,r}\lambda _{\hat{r}%
}^{r}\lambda _{\hat{\theta}}^{\theta }  \notag \\
& =\frac{3\mu \sin \alpha \sin \left( \phi -\Omega t\right) }{4m^{3}}  \notag
\\
& \times \left[ \ln \left( 1-\frac{2m}{r}\right) +\frac{m}{r}\left( \left( 1-%
\frac{2m}{r}\right) ^{-1}+1\right) \right]  \notag \\
& \times \left( 1-\frac{2m}{r}\right) ^{1/2}.  \label{B5}
\end{align}%
This is in agreement with \cite{RAM2001},\footnote{%
Our result is in agreement with their stationary vacuum magnetic field
external to a misaligned magnetized rotator, which is obtained as part of
the exterior solution of Maxwell's equations in the case of infinite
electrical conductivity.} and also in agreement with \cite{MY2012} and \cite%
{Sengupta1995} in the flat spacetime limit $m\rightarrow 0$ and in the
alignment limit $\alpha \rightarrow 0$, respectively. In Fig. \ref{fig3} are
shown the magnetic field lines based on Eqs. (\ref{B3})-(\ref{B5}), where we
set $\phi -\Omega t=0$.\footnote{%
To be precise, the actual field lines are drawn by means of Eq. (\ref{sp11}%
), where a collection of arbitrary constants chosen for the equation
correspond to magnetic field lines as shown in Fig. \ref{fig3}.} Note here
the difference between the field lines in curved spacetime and flat
spacetime: the blue curves represent the field lines in curved spacetime,
with the general relativistic effect taken into account through $m=GM/c^{2}$
for the star of mass $M$, while the green curves represent the field lines
in flat spacetime, in the limit $m\rightarrow 0$. The unity in the scale
used for this graph is equivalent to the radius of the neutron star, $%
R_{\ast }\simeq 10^{6}\,\mathrm{cm}$.

\section{Pulsar Emission and Effects of Obliquity\label{emission}}

\subsection{Radiative EM Fields and Pulsar Energetics\label{radiative}}

One of the most interesting features of an oblique rotator would be \textit{%
low-frequency} EM radiation produced by its obliquely rotating magnetic
dipole moment. We shall denote the magnetic dipole moment as $\mathbf{\mu }%
\left( t\right) $, a time--varying vector. The vector potential due to $%
\mathbf{\mu }\left( t\right) $, evaluated at the point $\mathbf{r}$ at time $%
t$ is (see \cite{MY2012})%
\begin{equation}
\mathbf{A}\left( t,\mathbf{r}\right) =\nabla \times \left( \frac{\mathbf{\mu 
}\left( t_{\mathrm{R}}\right) }{r}\right) ,  \label{rf1}
\end{equation}%
where $r=\left\vert \mathbf{r}\right\vert $ is the radial distance from the
centre of the rotator, and $t_{\mathrm{R}}\equiv t-r/c$ is the retarded
time. Expanding the right-hand side, we shall keep the term $\propto 1/r$,
which will survive and give a radially outward Poynting vector in the
far-field regime as will be shown below. This shall be termed the
\textquotedblleft radiative\textquotedblright\ potential \cite{MY2012}: 
\begin{equation}
\mathbf{A}\left( t,\mathbf{r}\right) \sim \frac{\mathbf{\dot{\mu}}\left( t_{%
\mathrm{R}}\right) \times \mathbf{e}_{\hat{r}}}{cr},  \label{rf2}
\end{equation}%
where $\mathbf{e}_{\hat{r}}=\mathbf{r}/\left\vert \mathbf{r}\right\vert $ is
the unit vector along the radial direction.\footnote{%
The way our radiative potential is calculated here resembles a
\textquotedblleft semi-relativistic treatment\textquotedblright\ in Ref. 
\cite{Ruffini1981}, in the following senses: (i)\ the field $\mathbf{A}%
\left( t,\mathbf{r}\right) $ radiates out in flat spacetime (to an observer
far away from the source), (ii) the source $\mathbf{\mu }\left( t_{\mathrm{R}%
}\right) $ contains the general relativistic information about its local
spacetime, which is strongly curved due to the mass of the neutron star.}

The magnetic dipole moment $\mathbf{\mu }$ is \textit{static} in the
corotating frame and always points along the $z^{\prime }$-axis:%
\begin{equation}
\mathbf{\mu }=\mu \,\mathbf{e}_{z^{\prime }},  \label{rf3}
\end{equation}%
where $\mu =\left\vert \mathbf{\mu }\right\vert $ is a constant. One can
convert this expression into spherical coordinates in the inertial frame,
which then becomes time--dependent:\footnote{%
The conversion can be achieved by means of Eqs. (\ref{e1})-(\ref{R}) in
Appendix \ref{appA}.} 
\begin{eqnarray}
\mathbf{\mu } &=&\mu \,\left\{ \left[ \cos \alpha \cos \theta +\sin \alpha
\sin \theta \cos \left( \phi -\Omega t\right) \right] \,\mathbf{e}_{\hat{r}%
}~_{_{_{{}}}}^{^{{}}}\right.  \notag \\
&&+\,\left[ -\cos \alpha \sin \theta +\sin \alpha \cos \theta \cos \left(
\phi -\Omega t\right) \right] \,\mathbf{e}_{\hat{\theta}}  \notag \\
&&\left. +\,\,\left[ -\sin \alpha \sin \left( \phi -\Omega t\right) \right]
\,\mathbf{e}_{\hat{\phi}}\right\} .  \label{rf4}
\end{eqnarray}%
Differentiation of Eq. (\ref{rf4}) with respect to time $t$ yields 
\begin{align}
\mathbf{\dot{\mu}}=& \Omega \mu \sin \alpha \left[ \sin \theta \sin \left(
\phi -\Omega t\right) \,\mathbf{e}_{\hat{r}}+\cos \theta \sin \left( \phi
-\Omega t\right) \,\mathbf{e}_{\hat{\theta}}~_{_{_{{}}}}^{^{{}}}\right. 
\notag \\
& \left. +\cos \left( \phi -\Omega t\right) \,\mathbf{e}_{\hat{\phi}}\right]
.  \label{rf5}
\end{align}

Substituting Eq. (\ref{rf5}) into Eq. (\ref{rf2}), we obtain the retarded
potential: 
\begin{eqnarray}
A_{\hat{\theta}} &=&\frac{\Omega \mu \sin \alpha \cos \left( \phi -\Omega t_{%
\mathrm{R}}\right) }{cr},  \label{rf6} \\
A_{\hat{\phi}} &=&-\frac{\Omega \mu \sin \alpha \cos \theta \sin \left( \phi
-\Omega t_{\mathrm{R}}\right) }{cr}.  \label{rf7}
\end{eqnarray}%
Then out of these the radiative EM\ fields can be finally determined: 
\begin{align}
B_{\hat{\theta}}& =-A_{\hat{\phi},r}=\frac{\Omega ^{2}\mu \sin \alpha \cos
\theta \cos \left( \phi -\Omega t_{\mathrm{R}}\right) }{c^{2}r}+\mathcal{O}%
\left( r^{-2}\right) ,  \label{rf8} \\
B_{\hat{\phi}}& =A_{\hat{\theta},r}=-\frac{\Omega ^{2}\mu \sin \alpha \sin
\left( \phi -\Omega t_{\mathrm{R}}\right) }{c^{2}r}+\mathcal{O}\left(
r^{-2}\right) ,  \label{rf9}
\end{align}%
and 
\begin{align}
E_{\hat{\theta}}& =-A_{\hat{\theta},t}c^{-1}=-\frac{\Omega ^{2}\mu \sin
\alpha \sin \left( \phi -\Omega t_{\mathrm{R}}\right) }{c^{2}r},
\label{rf10} \\
E_{\hat{\phi}}& =-A_{\hat{\phi},t}c^{-1}=-\frac{\Omega ^{2}\mu \sin \alpha
\cos \theta \cos \left( \phi -\Omega t_{\mathrm{R}}\right) }{c^{2}r},
\label{rf11}
\end{align}%
where the terms of $\mathcal{O}\left( r^{-2}\right) $ in Eqs. (\ref{rf8})
and (\ref{rf9}) can be disregarded in our analysis as they are too small to
consider in the far-field regime.

Now, with the EM fields given by Eqs. (\ref{rf8})-(\ref{rf11}) the Poynting
vector reads%
\begin{eqnarray}
S_{\hat{r}} &=&\frac{c}{4\pi }\left( E_{\hat{\theta}}B_{\hat{\phi}}-E_{\hat{%
\phi}}B_{\hat{\theta}}\right)  \notag \\
&=&\frac{\mu ^{2}\Omega ^{4}\sin ^{2}\alpha }{4\pi c^{3}r^{2}}  \notag \\
&&\times \left[ \sin ^{2}\left( \phi -\Omega t_{\mathrm{R}}\right) +\cos
^{2}\theta \cos ^{2}\left( \phi -\Omega t_{\mathrm{R}}\right) \right] .
\label{rf12}
\end{eqnarray}%
Then the power radiated by the EM fields can be evaluated as%
\begin{equation}
\mathcal{P}_{\mathrm{radiation}}=\int S_{\hat{r}}\,r^{2}d\varOmega=\frac{%
2\mu ^{2}\Omega ^{4}\sin ^{2}\alpha }{3c^{3}},  \label{rf13}
\end{equation}%
where $d\varOmega\equiv \sin \theta d\theta d\phi $ is a differential solid
angle. From this, it should be noted that the power $\mathcal{P}$ vanishes
in the alignment limit $\alpha \rightarrow 0$: it implies that the
\textquotedblleft obliqueness\textquotedblright\ of our rotating star is
responsible for radiation.

The above result can be applied to the well-known relation between pulsar
radiation and rotational energy loss. That is, the radiation power $\mathcal{%
P}_{\mathrm{radiation}}$ given by Eq. (\ref{rf13}) is equated to the rate of
loss of rotational kinetic energy $\mathcal{E}_{\mathrm{rotation}}$: 
\begin{equation}
\mathcal{P}_{\mathrm{radiation}}=-\,\mathcal{\dot{E}}_{\mathrm{rotation}},
\label{rf14}
\end{equation}%
where 
\begin{equation}
\mathcal{E}_{\mathrm{rotation}}=\frac{1}{2}J\Omega \simeq \frac{1}{5}%
\,MR_{\ast }^{2}\Omega ^{2},  \label{rf15}
\end{equation}%
due to Eq. (\ref{J}). In flat spacetime Eq. (\ref{rf14}) leads to 
\begin{equation}
\mu _{\mathrm{flat}}\simeq \frac{\sqrt{3}c^{3/2}M^{1/2}R_{\ast }}{2\sqrt{5}%
\pi \sin \alpha }\left( P\dot{P}\right) ^{1/2},  \label{rf16}
\end{equation}%
where $P=2\pi /\Omega $ is the rotational period of the star. However, in
curved spacetime, from Eq. (\ref{mu}) we may infer 
\begin{equation}
\mu _{\mathrm{curved}}=\left( 1-\frac{2m}{R_{\ast }}\right) ^{1/2}\mu _{%
\mathrm{flat}},  \label{rf17}
\end{equation}%
which implies $\mu _{\mathrm{curved}}\rightarrow \mu _{\mathrm{flat}}=\pi
R_{\ast }^{2}I$\ in the limit $m\rightarrow 0$ for the current loop model as
presented in Section \ref{Maxwell}. Then using Eqs. (\ref{rf16}) and (\ref%
{rf17}), we may express 
\begin{equation}
\mu _{\mathrm{curved}}\simeq \frac{\sqrt{3}c^{3/2}M^{1/2}R_{\ast }\left( 1-%
\frac{2m}{R_{\ast }}\right) ^{1/2}}{2\sqrt{5}\pi \sin \alpha }\left( P\dot{P}%
\right) ^{1/2}.  \label{rf17-1}
\end{equation}%
Now, by Eq. (\ref{B1}) we find the magnetic field strength at the polar cap
to be 
\begin{align}
B_{\ast }& \equiv B_{\hat{r}}\left( r=R_{\ast },\theta ^{\prime }=0\right) 
\notag \\
& =-\frac{3\mu _{\mathrm{curved}}}{4m^{3}}\left[ \ln \left( 1-\frac{2m}{%
R_{\ast }}\right) +\frac{2m}{R_{\ast }}\left( 1+\frac{m}{R_{\ast }}\right) %
\right] .  \label{rf18}
\end{align}%
Inserting Eq. (\ref{rf17-1}) into Eq. (\ref{rf18}), we finally obtain%
\begin{widetext}
\begin{equation}
B_{\ast \,\mathrm{curved}}\simeq -\frac{3\sqrt{3}\,c^{3/2}M^{1/2}R_{\ast }%
\left[ \ln \left( 1-\frac{2m}{R_{\ast }}\right) +\frac{2m}{R_{\ast }}\left(
1+\frac{m}{R_{\ast }}\right) \right] \left( 1-\frac{2m}{R_{\ast }}\right)
^{1/2}}{8\sqrt{5}\pi m^{3}\sin \alpha }\left( P\dot{P}\right) ^{1/2}.
\label{rf19}
\end{equation}
\end{widetext}This provides a general relativistic estimate of $B_{\ast }$
as $m=GM/c^{2}$ takes into account the effect of gravitation.

In the limit $m\rightarrow 0$, Eq. (\ref{rf19}) reduces to the expression
for $B_{\ast }$ in flat spacetime \cite{Carroll2007}: 
\begin{equation}
B_{\ast \,\mathrm{flat}}\simeq \frac{\sqrt{3}\,c^{3/2}M^{1/2}}{\sqrt{5}\pi
R_{\ast }^{2}\sin \alpha }\left( P\dot{P}\right) ^{1/2}.  \label{rf20}
\end{equation}%
\ 

Taking the ratio between the right-hand sides of Eqs. (\ref{rf19})\ and (\ref%
{rf20}), we find the general relativistic factor for the estimate of $%
B_{\ast }$:%
\begin{align}
& \mathrm{GR~factor}=\frac{B_{\ast \,\mathrm{curved}}}{B_{\ast \,\mathrm{flat%
}}}  \notag \\
& =-\frac{3R_{\ast }^{3}\left[ \ln \left( 1-\frac{2m}{R_{\ast }}\right) +%
\frac{2m}{R_{\ast }}\left( 1+\frac{m}{R_{\ast }}\right) \right] \left( 1-%
\frac{2m}{R_{\ast }}\right) ^{1/2}}{8m^{3}},  \label{rf21}
\end{align}%
which tends to $1$ in the flat spacetime limit $m\rightarrow 0$. For
example, the Crab pulsar (PSR B0531+21) has the radius $R_{\ast }\simeq
10^{6}\,\mathrm{cm}$ and $M\simeq 1.4\,M_{\odot }$, with observed data $%
P=0.0333\,\mathrm{s}$, $\dot{P}=4.21\times 10^{-13}$. Then $m=GM/c^{2}\simeq
2.065\times 10^{5}\,\mathrm{cm}$, and by Eqs. (\ref{rf19}), (\ref{rf20}) and
(\ref{rf21}) we have%
\begin{equation}
B_{\ast \,\mathrm{curved}}\sin \alpha \simeq 9.0\times 10^{12}\,\mathrm{G},
\label{rf22}
\end{equation}%
and 
\begin{equation}
B_{\ast \,\mathrm{flat}}\sin \alpha \simeq 8.0\times 10^{12}\,\mathrm{G},
\label{rf23}
\end{equation}%
with 
\begin{equation}
\text{GR factor}\simeq 1.12.  \label{rf24}
\end{equation}%
That is, the general relativistic effect increases the estimate of $B_{\ast
} $ by $12\,\%$. {\red The classical result (\ref{rf23}) is commonly used in
observational studies of various properties of the Crab pulsar; e.g., {in 
\cite{KT2015}} it is assumed that a field strength at the magnetic poles is
about $7.6\times 10^{12}\,\mathrm{G}$.} However, a correction due to the
GR-induced enhancement would not be significant as it keeps the estimate
still within the same order of magnitude. There are other things to take
into consideration, such as the dust and gas interacting with the pulsar in
the surrounding nebula, which would give rise to torques that contribute to
slowing down the pulsar's spin \cite{Carroll2007}. 

\subsection{Pulsar Radio Emission\label{radio}}

\subsubsection{Curvature Radiation\label{curv}}

As one of the possible mechanisms for pulsar radio emission, curvature
radiation can be discussed. It is emitted by a charged particle moving from
the region near the polar cap of our neutron star along a dipole magnetic
field line close to the magnetic axis, and received by an observer who is
far away from the star. The radiation can be obtained from the Li\'{e}%
nard--Wiechert potential. For a moving point charge $q$ at position $\mathbf{%
\xi }$, it is evaluated by an observer at the point $\mathbf{r}$ at time $t$
as: 
\begin{equation}
\mathbf{A}\left( t,\mathbf{r}\right) =\frac{q\mathbf{\dot{\xi}}\left( t_{%
\mathrm{R}}\right) }{c\left( 1-\frac{\mathbf{\dot{\xi}}\left( t_{\mathrm{R}%
}\right) }{c}\cdot \mathbf{n}\right) \left\vert \mathbf{r-\xi }\left( t_{%
\mathrm{R}}\right) \right\vert },  \label{lw}
\end{equation}%
where $\mathbf{n}=\left( \mathbf{r-\xi }\left( t_{\mathrm{R}}\right) \right)
/\left\vert \mathbf{r-\xi }\left( t_{\mathrm{R}}\right) \right\vert $ is the
unit vector pointing in the direction from the charge to the field point,
and $t_{\mathrm{R}}\equiv t-\left\vert \mathbf{r-\xi }\left( t_{\mathrm{R}%
}\right) \right\vert /c$ is the retarded time, and the overdot denotes
differentiation with respect to $t_{\mathrm{R}}$. A simple configuration for
this emission mechanism is illustrated in Fig. \ref{fig3}: the observer's
line of sight, i.e., the line connecting the two points $\mathbf{\xi }$ and $%
\mathbf{r}$ is represented by the purple dashed line.

In this mechanism, a charge is assumed to move along a magnetic field line;
therefore, its trajectory $\mathbf{\xi }$ is identified with the field line.
Then its velocity is tangent to the field line at the point $\mathbf{\xi }$
and can be written as 
\begin{equation}
\mathbf{\dot{\xi}}=\dot{\xi}\,\mathbf{\hat{v}}=\beta c\mathbf{\hat{B}}\left( 
\mathbf{\xi }\right) ,  \label{sp1}
\end{equation}%
where $\dot{\xi}\,=\beta c=$ $\mathrm{const.}$, and $\mathbf{\hat{v}}\equiv 
\mathbf{\hat{B}}\left( \mathbf{\xi }\right) =\mathbf{B}\left( \mathbf{\xi }%
\right) /\left\vert \mathbf{B}\left( \mathbf{\xi }\right) \right\vert $
denotes the unit vector along which the radiation pulse propagates: it also
coincides with $\mathbf{n}=\left( \mathbf{r-\xi }\right) /\left\vert \mathbf{%
r-\xi }\right\vert $, the unit vector along the line of sight.

In the same plane where a magnetic field line belongs, the acceleration of a
charge that moves along the field line can be expressed as%
\begin{equation}
\mathbf{\ddot{\xi}}=\beta c\,\dot{\hat{\mathbf{v}}}=\beta c\dot{\hat{\mathbf{%
B}}}(\mathbf{\xi })=\ddot{\xi}\mathbf{\hat{a}}=\beta c\left\vert \dot{\hat{%
\mathbf{B}}}(\mathbf{\xi })\right\vert \left( \dot{\hat{\mathbf{B}}}(\mathbf{%
\xi })/\left\vert \dot{\hat{\mathbf{B}}}(\mathbf{\xi })\right\vert \right) ,
\label{sp2}
\end{equation}%
where $\ddot{\xi}=\beta c\left\vert \dot{\hat{\mathbf{B}}}(\mathbf{\xi }%
)\right\vert $, and $\mathbf{\hat{a}}\equiv \dot{\hat{\mathbf{B}}}(\mathbf{%
\xi })/\left\vert \dot{\hat{\mathbf{B}}}(\mathbf{\xi })\right\vert $ denotes
the unit vector along which the radiation pulse is polarized. From $\dot{\xi}%
\,=\beta c=$ $\mathrm{const.}$, one can easily check the orthogonality, $%
\mathbf{\dot{\xi}\cdot \ddot{\xi}}=0$.

Now, in the corotating frame, magnetic field lines are static, and therefore
the description of motion of a charge along a field line would be much
simpler than in the inertial frame. By means of Eqs. (\ref{B1}), (\ref{B2}),
(\ref{sp1}) and (\ref{sp2}) the velocity and the acceleration are expressed
respectively as

\begin{equation}
\mathbf{\dot{\xi}}=\beta c\mathbf{\hat{v}};\mathrm{\ with}\mathbf{~\hat{v}}=%
\mathbf{\,}\frac{B_{\hat{r}}\left( \mathbf{\xi }\right) \mathbf{e}_{\hat{r}%
}+B_{\hat{\theta}^{\prime }}\left( \mathbf{\xi }\right) \,\mathbf{e}_{\hat{%
\theta}^{\prime }}}{\sqrt{B_{\hat{r}}^{2}\left( \mathbf{\xi }\right) +B_{%
\hat{\theta}^{\prime }}^{2}\left( \mathbf{\xi }\right) }},  \label{sp3}
\end{equation}%
and

\begin{equation}
\mathbf{\ddot{\xi}}=\frac{\beta ^{2}c^{2}}{\rho }\mathbf{\hat{a}};\mathrm{\
with}\mathbf{~\hat{a}}=\frac{-B_{\hat{\theta}^{\prime }}\left( \mathbf{\xi }%
\right) \mathbf{e}_{\hat{r}}+\,B_{\hat{r}}\left( \mathbf{\xi }\right) 
\mathbf{e}_{\hat{\theta}^{\prime }}}{\sqrt{B_{\hat{r}}^{2}\left( \mathbf{\xi 
}\right) +B_{\hat{\theta}^{\prime }}^{2}\left( \mathbf{\xi }\right) }},
\label{sp4}
\end{equation}%
where $\rho $ is the curvature radius of the charge's trajectory along a
magnetic field line, which is defined through Eqs. (\ref{sp8})-(\ref{sp13})
below.

The trajectory of a source charge moving along a magnetic field line can be
expressed by 
\begin{equation}
\mathbf{\xi }_{\mathrm{s}}=r_{\mathrm{s}}\,\mathbf{e}_{\hat{r}}\text{ }%
\mathrm{at}\text{ }\theta _{\mathrm{s}}^{\prime },  \label{sp5}
\end{equation}%
where the subscript \textquotedblleft $\mathrm{s}$\textquotedblright\
denotes the source charge: the trajectory is equivalent to the field line
given by Eqs. (\ref{B1}) and (\ref{B2}) defined at the location of the
source charge $\left( r_{\mathrm{s}},\theta _{\mathrm{s}}^{\prime }\right) $
in the corotating frame. Further, the velocity and the acceleration of the
source charge can be expressed using (\ref{sp3}) and (\ref{sp4}): by means
of Eqs. (\ref{B1}) and (\ref{B2}), we have 
\begin{equation}
\mathbf{\dot{\xi}}_{\mathrm{s}}=\beta c\mathbf{\hat{v}};\mathrm{\ with}%
\mathbf{~\hat{v}}=\frac{-g\left( r_{\mathrm{s}}\right) \,\mathbf{e}_{\hat{r}%
}+\tan \theta _{\mathrm{s}}^{\prime }\,\mathbf{e}_{\hat{\theta}^{\prime }}}{%
\sqrt{g^{2}\left( r_{\mathrm{s}}\right) +\tan ^{2}\theta _{\mathrm{s}%
}^{\prime }}},  \label{sp6}
\end{equation}%
and

\begin{equation}
\mathbf{\ddot{\xi}}_{\mathrm{s}}=\frac{\beta ^{2}c^{2}}{\rho }\mathbf{\hat{a}%
};\mathrm{\ with}\mathbf{~\hat{a}}=\frac{-\tan \theta _{\mathrm{s}}^{\prime
}\,\mathbf{e}_{\hat{r}}-g\left( r_{\mathrm{s}}\right) \mathbf{e}_{\hat{\theta%
}^{\prime }}}{\sqrt{g^{2}\left( r_{\mathrm{s}}\right) +\tan ^{2}\theta _{%
\mathrm{s}}^{\prime }}},  \label{sp7}
\end{equation}%
where%
\begin{align}
& g\left( r\right)  \notag \\
& \equiv \frac{\ln \left( 1-\frac{2m}{r}\right) +\frac{2m}{r}\left( 1+\frac{m%
}{r}\right) }{\left[ \ln \left( 1-\frac{2m}{r}\right) +\frac{m}{r}\left(
\left( 1-\frac{2m}{r}\right) ^{-1}+1\right) \right] \left( 1-\frac{2m}{r}%
\right) ^{1/2}}.  \label{dv9}
\end{align}

Viewed in the inertial frame, $\mathbf{\hat{v}}$ and $\mathbf{\hat{a}}$ in
Eqs. (\ref{sp6}) and (\ref{sp7}) can be set to lie in the $\hat{r}\hat{\theta%
}$-plane. A magnetic field line can be confined to this plane via Eq. (\ref%
{e4}) and by defining Eqs. (\ref{B3})-(\ref{B5}) at $\mathbf{\xi }_{\mathrm{s%
}}=\left( r_{\mathrm{s}},\theta _{\mathrm{s}},\phi _{\mathrm{s}}\right) $
with $\phi _{\mathrm{s}}=\Omega t_{\mathrm{R}}$. Further, we have the
magnetic colatitude being reduced to $\theta _{\mathrm{s}}^{\prime }=\theta
_{\mathrm{s}}-\alpha $ in the plane via Eq. (\ref{col}). In view of Fig. \ref%
{fig3}, it should be noted that the trajectory $\mathbf{\xi }_{\mathrm{s}}$,
the velocity $\mathbf{\dot{\xi}}_{\mathrm{s}}$ and the acceleration $\mathbf{%
\ddot{\xi}}_{\mathrm{s}}$ of a source charge may describe its motion either
in curved spacetime or in flat spacetime. That is, the charge moves along
the magnetic field lines in blue in curved spacetime, with the general
relativistic effect taken into account in Eqs. (\ref{B3})-(\ref{B5}) through 
$m=GM/c^{2}$ for a neutron star of mass $M$, while it moves along the
magnetic field lines in green in flat spacetime, in the limit $m\rightarrow
0 $. Notable differences between the cases of curved and flat spacetimes are
the directions of $\mathbf{\dot{\xi}}_{\mathrm{s}}$\ and $\mathbf{\ddot{\xi}}%
_{\mathrm{s}}$, which can be easily checked from Eqs. (\ref{sp6}) and (\ref%
{sp7}). However, the most significant distinction between the two cases is
characterized by $\rho $, the curvature radius of the charge's trajectory:
the general relativistic effect, which is due to strong gravity in the
pulsar magnetosphere, causes the curvature radius to shrink and thence the
acceleration to increase since $\left\vert \mathbf{\ddot{\xi}}_{\mathrm{s}%
}\right\vert =\beta ^{2}c^{2}/\rho $ according to Eq. (\ref{sp7}).\footnote{%
In contrast to this, the general relativistic effect causes the direction of
the velocity to change but not its magnitude, $\left\vert \mathbf{\dot{\xi}}%
\right\vert =\beta c$ according to Eq. (\ref{sp6}).}

The curvature radius $\rho $ in the general relativistic context can be
defined from an infinitesimal path of a charge moving along a magnetic field
line in curved spacetime. Considering the charge's trajectory as described
by Eq. (\ref{sp5}), one can write down the infinitesimal path:%
\begin{eqnarray}
d\mathbf{l} &=&d\hat{r}_{\mathrm{s}}\,\mathbf{e}_{\hat{r}}+d\hat{\theta}_{%
\mathrm{s}}^{\prime }\,\mathbf{e}_{\hat{\theta}^{\prime }}  \notag \\
&=&\left( 1-\frac{2m}{r_{\mathrm{s}}}\right) ^{-1/2}dr_{\mathrm{s}}\,\mathbf{%
e}_{\hat{r}}+r_{\mathrm{s}}d\theta _{\mathrm{s}}^{\prime }\,\mathbf{e}_{\hat{%
\theta}^{\prime }}.  \label{sp8}
\end{eqnarray}%
This is defined in a local Lorentz frame by means of the tetrad (\ref{te});
that is, with $d\hat{r}=dr\left( \lambda _{\hat{r}}^{r}\right) ^{-1}$ and $d%
\hat{\theta}^{\prime }=d\theta ^{\prime }\left( \lambda _{\hat{\theta}%
^{\prime }}^{\theta ^{\prime }}\right) ^{-1}$ being defined at $\mathbf{\xi }%
_{\mathrm{s}}=\left( r_{\mathrm{s}},\theta _{\mathrm{s}}^{\prime }\right) $.
The infinitesimal arc length along the field line is then given by%
\begin{align}
& dl=\left\vert d\mathbf{l}\right\vert  \notag \\
& =r_{\mathrm{s}}d\theta _{\mathrm{s}}^{\prime }\left[ 1+r_{\mathrm{s}%
}^{-2}\left( 1-\frac{2m}{r_{\mathrm{s}}}\right) ^{-1}\left( \frac{%
B_{r}\left( \mathbf{\xi }_{\mathrm{s}}\right) \,}{B_{\theta ^{\prime
}}\left( \mathbf{\xi }_{\mathrm{s}}\right) }\right) ^{2}\right] ^{1/2}.
\label{sp9}
\end{align}%
Here the second term inside the square brackets is due to the differential
equation \cite{KK2000,DKL2000}: 
\begin{equation}
~\frac{dr_{\mathrm{s}}}{d\theta _{\mathrm{s}}^{\prime }}=\frac{B_{r}\left( 
\mathbf{\xi }_{\mathrm{s}}\right) \,}{B_{\theta ^{\prime }}\left( \mathbf{%
\xi }_{\mathrm{s}}\right) },  \label{sp10}
\end{equation}%
where $B_{r}\left( \mathbf{\xi }_{\mathrm{s}}\right) $ and $B_{\theta
^{\prime }}\left( \mathbf{\xi }_{\mathrm{s}}\right) $ mean $B_{\hat{r}%
}\left( \lambda _{\hat{r}}^{r}\right) ^{-1}$ and $B_{\hat{\theta}^{\prime
}}\left( \lambda _{\hat{\theta}^{\prime }}^{\theta ^{\prime }}\right) ^{-1}$
defined at $\mathbf{\xi }_{\mathrm{s}}=\left( r_{\mathrm{s}},\theta _{%
\mathrm{s}}^{\prime }\right) $, respectively, by means of Eqs. (\ref{B1}), (%
\ref{B2}) and (\ref{te}). A solution to Eq. (\ref{sp10}) is a curve 
\begin{equation}
A_{\phi ^{\prime }}\left( \mathbf{\xi }_{\mathrm{s}}\right) =\mathrm{const.},
\label{sp11}
\end{equation}%
where $A_{\phi ^{\prime }}\left( \mathbf{\xi }_{\mathrm{s}}\right) $ is
given by Eq. (\ref{A1}) defined at $\mathbf{\xi }_{\mathrm{s}}=\left( r_{%
\mathrm{s}},\theta _{\mathrm{s}}^{\prime }\right) $.\footnote{%
The solution can easily be verified: $A_{\phi ^{\prime },r}\left( \mathbf{%
\xi }_{\mathrm{s}}\right) \,dr_{\mathrm{s}}+A_{\phi ^{\prime },\theta
^{\prime }}\left( \mathbf{\xi }_{\mathrm{s}}\right) \,d\theta _{\mathrm{s}%
}^{\prime }=0$ $\Leftrightarrow $ $-B_{\theta ^{\prime }}\left( \mathbf{\xi }%
_{\mathrm{s}}\right) \,dr_{\mathrm{s}}+B_{r}\left( \mathbf{\xi }_{\mathrm{s}%
}\right) \,d\theta _{\mathrm{s}}^{\prime }=0$.} Out of Eq. (\ref{sp9}) one
can finally define the curvature radius of a charge's trajectory along a
magnetic field line \cite{KK2000}: 
\begin{equation}
\rho \equiv \frac{dl}{d\theta _{\mathrm{s}}^{\prime }}=r_{\mathrm{s}}\left[
1+\left( \frac{B_{\hat{r}}\left( \mathbf{\xi }_{\mathrm{s}}\right) }{B_{\hat{%
\theta}^{\prime }}\left( \mathbf{\xi }_{\mathrm{s}}\right) }\right) ^{2}%
\right] ^{1/2}.  \label{sp13}
\end{equation}

Having Eqs. (\ref{B1}) and (\ref{B2}) defined at $\mathbf{\xi }_{\mathrm{s}%
}=\left( r_{\mathrm{s}},\theta _{\mathrm{s}}^{\prime }\right) $, the
curvature radius is evaluated from Eq. (\ref{sp13}): \ 
\begin{equation}
\rho =r_{\mathrm{s}}\left[ 1+\,g^{2}\left( r_{\mathrm{s}}\right) \cot
^{2}\theta _{\mathrm{s}}^{\prime }\right] ^{1/2},  \label{sp14}
\end{equation}%
where $g\left( r\right) $ refers to Eq. (\ref{dv9}). In the flat spacetime
limit $m\rightarrow 0$, we have $g\left( r_{\mathrm{s}}\right) \rightarrow
-2 $, and thence 
\begin{equation}
\rho =r_{\mathrm{s}}\left[ 1+\,4\cot ^{2}\theta _{\mathrm{s}}^{\prime }%
\right] ^{1/2}.  \label{sp15}
\end{equation}%
Obviously, the difference between Eqs. (\ref{sp14}) and (\ref{sp15}) shows
the general relativistic effect due to gravity in the pulsar magnetosphere: $%
\rho _{\mathrm{curved}}\leq \rho _{\mathrm{flat}}$ as $g^{2}\left( r_{%
\mathrm{s}}\right) \leq 4$.\footnote{%
For example, for curvature radiation from the pulsar PSR J1828-1101 (see
Section \ref{profile}), the emission heights of source charges are estimated
to be around $35$ to $45\,\mathrm{km}$, relative to the center of the
neutron star \cite{JK2019}. Then for $\theta _{\mathrm{s}}^{\prime }\ll 1$,
we have the ratio $\rho _{\mathrm{curved}}/\rho _{\mathrm{flat}}\approx $ $%
\left\vert g\left( r_{\mathrm{s}}\right) \right\vert /2\approx 0.96$ to $%
0.97 $, where $r_{\mathrm{s}}\simeq 3.5R_{\ast }$ to $4.5R_{\ast }\simeq 14m$
to $18m$.}

From Eq. (\ref{sp1}) and the definition of the curvature radius, we find%
\begin{equation}
\mathbf{\dot{\xi}}_{\mathrm{s}}=\beta c\mathbf{\hat{v}}=\dot{l}\,\mathbf{%
\hat{v}}=\rho \dot{\theta}_{\mathrm{s}}^{\prime }\,\mathbf{\hat{v}}.
\label{sp16}
\end{equation}%
From this the frequency of radiation by a moving charge is obtained: 
\begin{equation}
\omega _{\mathrm{o}}\equiv \dot{\theta}_{\mathrm{s}}^{\prime }=\frac{\beta c%
}{\rho }.  \label{sp17}
\end{equation}%
However, for a source charge moving near the speed of light, i.e., $\gamma
=\left( 1-\beta ^{2}\right) ^{-1/2}\gg 1$, we take into consideration the 
\textit{relativistic beaming} effect, and the characteristic frequency for
the curvature radiation should be expressed as \cite{RS1975} 
\begin{equation}
\omega _{\mathrm{c}}=\frac{3}{2}\gamma ^{3}\frac{\beta c}{\rho }.
\label{sp18}
\end{equation}

Out of the potential (\ref{lw}), the electric field is derived: 
\begin{eqnarray}
\mathbf{E}\left( t,\mathbf{r}\right) &=&\frac{q\mathbf{n}\times \left[
\left( \mathbf{n}-\frac{\mathbf{\dot{\xi}}\left( t_{\mathrm{R}}\right) }{c}%
\right) \times \frac{\mathbf{\ddot{\xi}}\left( t_{\mathrm{R}}\right) }{c}%
\right] }{c\left( 1-\frac{\mathbf{\dot{\xi}}\left( t_{\mathrm{R}}\right) }{c}%
\cdot \mathbf{n}\right) ^{3}\left\vert \mathbf{r-\xi }\left( t_{\mathrm{R}%
}\right) \right\vert }  \notag \\
&&+\frac{q\mathbf{n}\left( \mathbf{n}-\frac{\mathbf{\dot{\xi}}\left( t_{%
\mathrm{R}}\right) }{c}\right) }{\gamma ^{2}\left( 1-\frac{\mathbf{\dot{\xi}}%
\left( t_{\mathrm{R}}\right) }{c}\cdot \mathbf{n}\right) ^{3}\left\vert 
\mathbf{r-\xi }\left( t_{\mathrm{R}}\right) \right\vert ^{2}}.  \label{sp19}
\end{eqnarray}%
Here the first term $\sim \left\vert \mathbf{r-\xi }\left( t_{\mathrm{R}%
}\right) \right\vert ^{-1}$ pertains to \textit{radiation} by a moving
charge, requiring the charge acceleration $\mathbf{\ddot{\xi}}$, whereas the
second term $\sim \left\vert \mathbf{r-\xi }\left( t_{\mathrm{R}}\right)
\right\vert ^{-2}$ refers to the \textit{static} part of the electric field
of the charge and will be disregarded in our analysis.

For a distant observer, i.e., $\left\vert \mathbf{r}\right\vert \gg
\left\vert \mathbf{\xi }_{\mathrm{s}}\right\vert $, we have $\mathbf{n}%
=\left( \mathbf{r-\xi }_{\mathrm{s}}\right) /\left\vert \mathbf{r-\xi }_{%
\mathrm{s}}\right\vert \approx \mathbf{\hat{r}}=\mathbf{r}/\left\vert 
\mathbf{r}\right\vert $, and therefore from (\ref{sp19}) the electric field
of radiation can be expressed as \cite{Pacholczyk1970,Jackson1976,GS1990} 
\begin{equation}
\mathbf{E}\left( t,\mathbf{r}\right) \approx \frac{q\mathbf{\hat{r}}\times %
\left[ \left( \mathbf{\hat{r}}-\frac{\mathbf{\dot{\xi}}_{\mathrm{s}}\left(
t_{\mathrm{R}}\right) }{c}\right) \times \frac{\mathbf{\ddot{\xi}}_{\mathrm{s%
}}\left( t_{\mathrm{R}}\right) }{c}\right] }{c\left( 1-\frac{\mathbf{\dot{\xi%
}}_{\mathrm{s}}\left( t_{\mathrm{R}}\right) }{c}\cdot \mathbf{\hat{r}}%
\right) ^{3}\left\vert \mathbf{r-\xi }_{\mathrm{s}}\left( t_{\mathrm{R}%
}\right) \right\vert },  \label{sp20}
\end{equation}%
where%
\begin{align}
\mathbf{\xi }_{\mathrm{s}}& =\rho \left( \sin \vartheta ,0,\cos \vartheta
\right) ,  \label{sp20-1} \\
\mathbf{\dot{\xi}}_{\mathrm{s}}& =\beta c\left( \cos \vartheta ,0,-\sin
\vartheta \right) ,  \label{sp21} \\
\mathbf{\ddot{\xi}}_{\mathrm{s}}& =\frac{\beta ^{2}c^{2}}{\rho }\left( -\sin
\vartheta ,0,-\cos \vartheta \right) ,  \label{sp22}
\end{align}%
refer to the trajectory, velocity and acceleration of a source charge,
respectively in the corotating frame given by (\ref{sp5})-(\ref{sp7}), being
expressed in a Cartesian frame specially chosen for computational
convenience, and%
\begin{equation}
\mathbf{\hat{r}}=\left( \cos \varphi ,\sin \varphi ,0\right) .  \label{sp23}
\end{equation}%
Here the motion of the charge along the field line is parametrized by $t_{%
\mathrm{R}}$ via the polar angle $\vartheta $, i.e., $\vartheta =$ $\beta
ct_{\mathrm{R}}/\rho $, whereas the observational direction is parametrized
by the azimuthal angle $\varphi $. A simple geometrical configuration for
curvature radiation viewed in this Cartesian frame is illustrated in Fig. %
\ref{fig4}. In Appendix \ref{appB} the technical details are provided
regarding how the Cartesian expressions for $\mathbf{\xi }_{\mathrm{s}}$, $%
\mathbf{\dot{\xi}}_{\mathrm{s}}$ and $\mathbf{\ddot{\xi}}_{\mathrm{s}}$
above can be obtained through the coordinate transformations of the initial
spherical polar representations (\ref{sp5})-(\ref{sp7}).

\begin{figure}[tbh]
\centering\includegraphics[width=8.6cm]{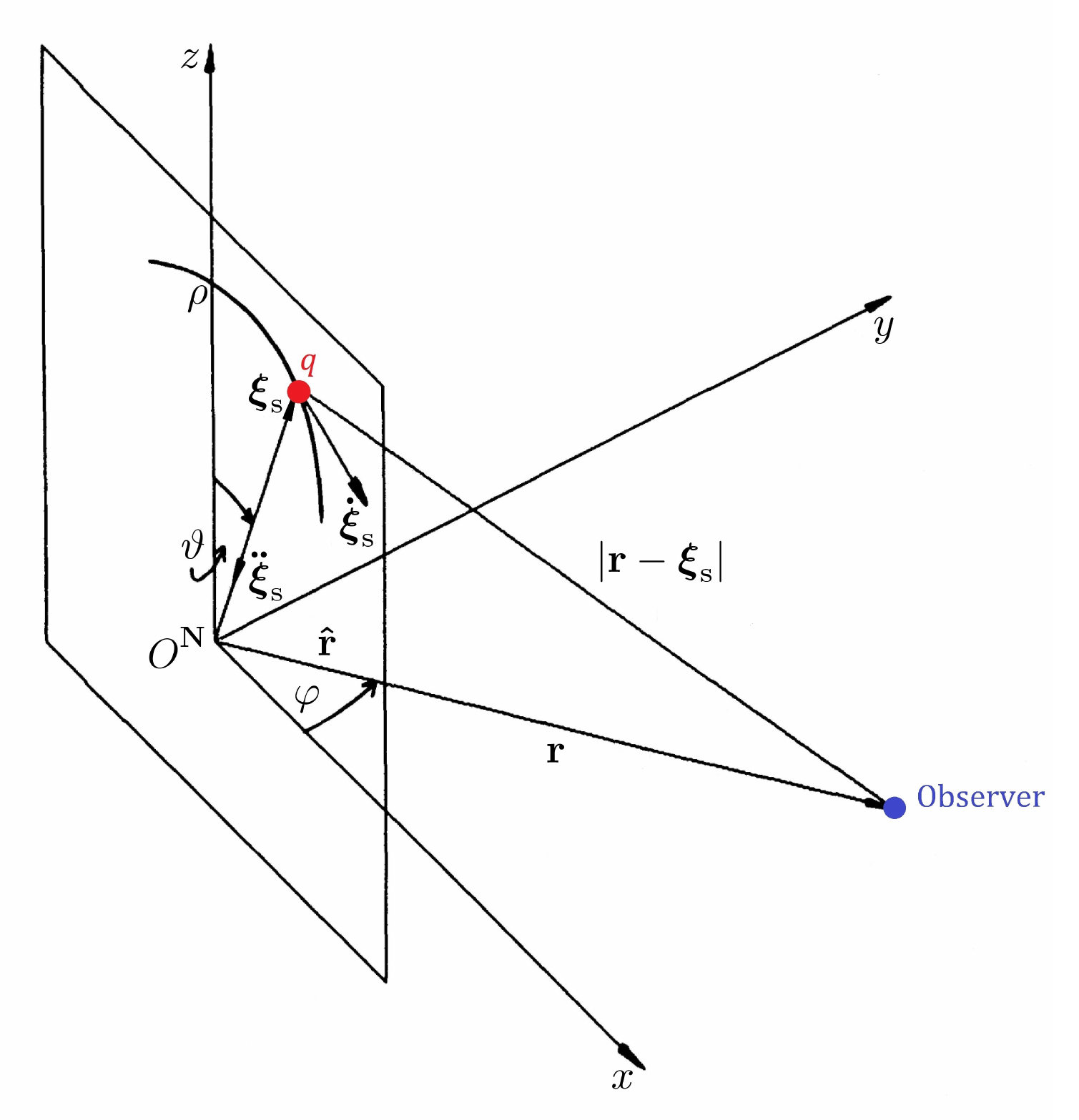}
\caption{A geometrical configuration for curvature radiation viewed in a
specially chosen Cartesian frame. Note that the frame is centred at a new
origin $O^{\mathrm{N}}$ (see Appendix \protect\ref{appB} for technical
details). (Credit: Gil and Snakowski \protect\cite{GS1990}; reproduced with
some modifications.)}
\label{fig4}
\end{figure}

\subsubsection{Pulse Profiles\label{profile}}

Following \cite{Pacholczyk1970,Jackson1976,GS1990,RL1979}, one can express
Stokes parameters out of the radiation field (\ref{sp20}), which describe
polarization properties of the curvature radiation discussed above: 
\begin{align}
& I=\tilde{E}_{\parallel }^{\ast }\tilde{E}_{\parallel }+\tilde{E}_{\perp
}^{\ast }\tilde{E}_{\perp }  \notag \\
& =\mathcal{E}_{\mathrm{o}}^{2}\omega ^{2}\left[ \left( \delta ^{2}+\varphi
^{2}\right) ^{2}\mathrm{K}_{2/3}^{2}\left( \frac{\omega }{3\omega _{\mathrm{o%
}}}\left( \delta ^{2}+\varphi ^{2}\right) ^{3/2}\right) \right.  \notag \\
& \left. +\,\varphi ^{2}\left( \delta ^{2}+\varphi ^{2}\right) ^{2}\mathrm{K}%
_{1/3}^{2}\left( \frac{\omega }{3\omega _{\mathrm{o}}}\left( \delta
^{2}+\varphi ^{2}\right) ^{3/2}\right) \right] ,  \label{pp1} \\
& Q=\tilde{E}_{\parallel }^{\ast }\tilde{E}_{\parallel }-\tilde{E}_{\perp
}^{\ast }\tilde{E}_{\perp }  \notag \\
& =\mathcal{E}_{\mathrm{o}}^{2}\omega ^{2}\left[ \left( \delta ^{2}+\varphi
^{2}\right) ^{2}\mathrm{K}_{2/3}^{2}\left( \frac{\omega }{3\omega _{\mathrm{o%
}}}\left( \delta ^{2}+\varphi ^{2}\right) ^{3/2}\right) \right.  \notag \\
& \left. -\,\varphi ^{2}\left( \delta ^{2}+\varphi ^{2}\right) ^{2}\mathrm{K}%
_{1/3}^{2}\left( \frac{\omega }{3\omega _{\mathrm{o}}}\left( \delta
^{2}+\varphi ^{2}\right) ^{3/2}\right) \right] ,  \label{pp2} \\
& U=\tilde{E}_{\parallel }^{\ast }\tilde{E}_{\perp }+\tilde{E}_{\parallel }%
\tilde{E}_{\perp }^{\ast }  \notag \\
& =2\cos \psi \mathcal{E}_{\mathrm{o}}^{2}\omega ^{2}\varphi \left( \delta
^{2}+\varphi ^{2}\right) ^{2}  \notag \\
& \times \mathrm{K}_{2/3}\left( \frac{\omega }{3\omega _{\mathrm{o}}}\left(
\delta ^{2}+\varphi ^{2}\right) ^{3/2}\right) \mathrm{K}_{1/3}\left( \frac{%
\omega }{3\omega _{\mathrm{o}}}\left( \delta ^{2}+\varphi ^{2}\right)
^{3/2}\right) ,  \label{pp3} \\
& V=-\mathrm{i}\left( \tilde{E}_{\parallel }^{\ast }\tilde{E}_{\perp }-%
\tilde{E}_{\parallel }\tilde{E}_{\perp }^{\ast }\right)  \notag \\
& =2\sin \psi \mathcal{E}_{\mathrm{o}}^{2}\omega ^{2}\varphi \left( \delta
^{2}+\varphi ^{2}\right) ^{2}  \notag \\
& \times \mathrm{K}_{2/3}\left( \frac{\omega }{3\omega _{\mathrm{o}}}\left(
\delta ^{2}+\varphi ^{2}\right) ^{3/2}\right) \mathrm{K}_{1/3}\left( \frac{%
\omega }{3\omega _{\mathrm{o}}}\left( \delta ^{2}+\varphi ^{2}\right)
^{3/2}\right) ,  \label{pp4}
\end{align}%
where $\tilde{E}_{\parallel }$ and $\tilde{E}_{\perp }$ denote the Fourier
transform $\mathbf{\tilde{E}}\left( \omega \right) =\int_{-\infty }^{\infty }%
\mathbf{E}\left( t\right) \exp \left( \mathrm{i}\omega t\right) \,dt$,
decomposed as $\tilde{E}_{\parallel }\left( \omega \right) \,\mathbf{e}_{z}+%
\tilde{E}_{\perp }\left( \omega \right) \,\mathbf{e}_{y}$ in the Cartesian
frame as illustrated in Fig. \ref{fig4}, and $^{\ast }$ means the complex
conjugate, and $\mathcal{E}_{\mathrm{o}}=q\omega _{\mathrm{o}}/\left( 2\sqrt{%
3}\pi ^{2}cr\right) $ with $\omega _{\mathrm{o}}$ being defined by Eq. (\ref%
{sp17}), and $\delta \equiv \gamma ^{-1}=\left( 1-\beta ^{2}\right)
^{1/2}\ll 1$ is the half-angle of the beam emission, and $\mathrm{K}_{1/3}$
and $\mathrm{K}_{2/3}$ denote the modified Bessel functions of the second
kind, and $\psi =\psi _{\parallel }-\psi _{\perp }$ is the phase difference
between the two components of the electric field having different initial
phases; $E_{\parallel }\left( t\right) \sim \exp \left[ \mathrm{i}\left(
\omega t+\psi _{\parallel }\right) \right] $ and $E_{\perp }\left( t\right)
\sim \exp \left[ \mathrm{i}\left( \omega t+\psi _{\perp }\right) \right] $.
With regard to the polarization state, $I$ is a measure of the total
intensity, $Q$ and $U$ jointly describe the linear polarization, and $V$
describes the circular polarization of the curvature radiation. We plot
these parameters as functions of the phase angle $\varphi $ to simulate the
pulse profiles of pulsar radio emission theoretically.

Apart from the characteristic frequency $\omega _{\mathrm{c}}$ given by (\ref%
{sp18}), one can define the \textit{peak} frequency $\omega _{\mathrm{p}}$
of the intensity $I$ at the center of the beam \cite{GS1990}: 
\begin{equation}
\left. \frac{\partial I\left( \omega ,\varphi \right) }{\partial \omega }%
\right\vert _{\varphi =0,\omega =\omega _{\mathrm{p}}}=0.  \label{pp5}
\end{equation}%
Then using Eq. (\ref{pp1}), it turns out 
\begin{equation}
\omega _{\mathrm{p}}\approx 1.251\gamma ^{3}\omega _{\mathrm{o}}=1.251\gamma
^{3}\frac{\beta c}{\rho },  \label{pp6}
\end{equation}%
which is of the same order as the characteristic frequency $\omega _{\mathrm{%
c}}$ given by (\ref{sp18}). However, at $\varphi =0$, the argument of the
modified Bessel functions in (\ref{pp1})-(\ref{pp4}) above becomes 
\begin{equation}
\frac{\delta ^{3}\omega }{3\omega _{\mathrm{o}}}=\frac{\omega }{3\gamma
^{3}\omega _{\mathrm{o}}}\approx 0.417\frac{\omega }{\omega _{\mathrm{p}}}.
\label{pp7}
\end{equation}%
Hence, the ratio $\omega /\omega _{\mathrm{p}}\approx 1.2\omega /\omega _{%
\mathrm{c}}$ will play an important role in determining the general features
of the pulse profiles of the Stokes parameters plotted from (\ref{pp1})-(\ref%
{pp4}) above.

\begin{figure*}[tbp]
\subfloat[Actual data plots]{
\includegraphics[angle=0, width=9.4cm]{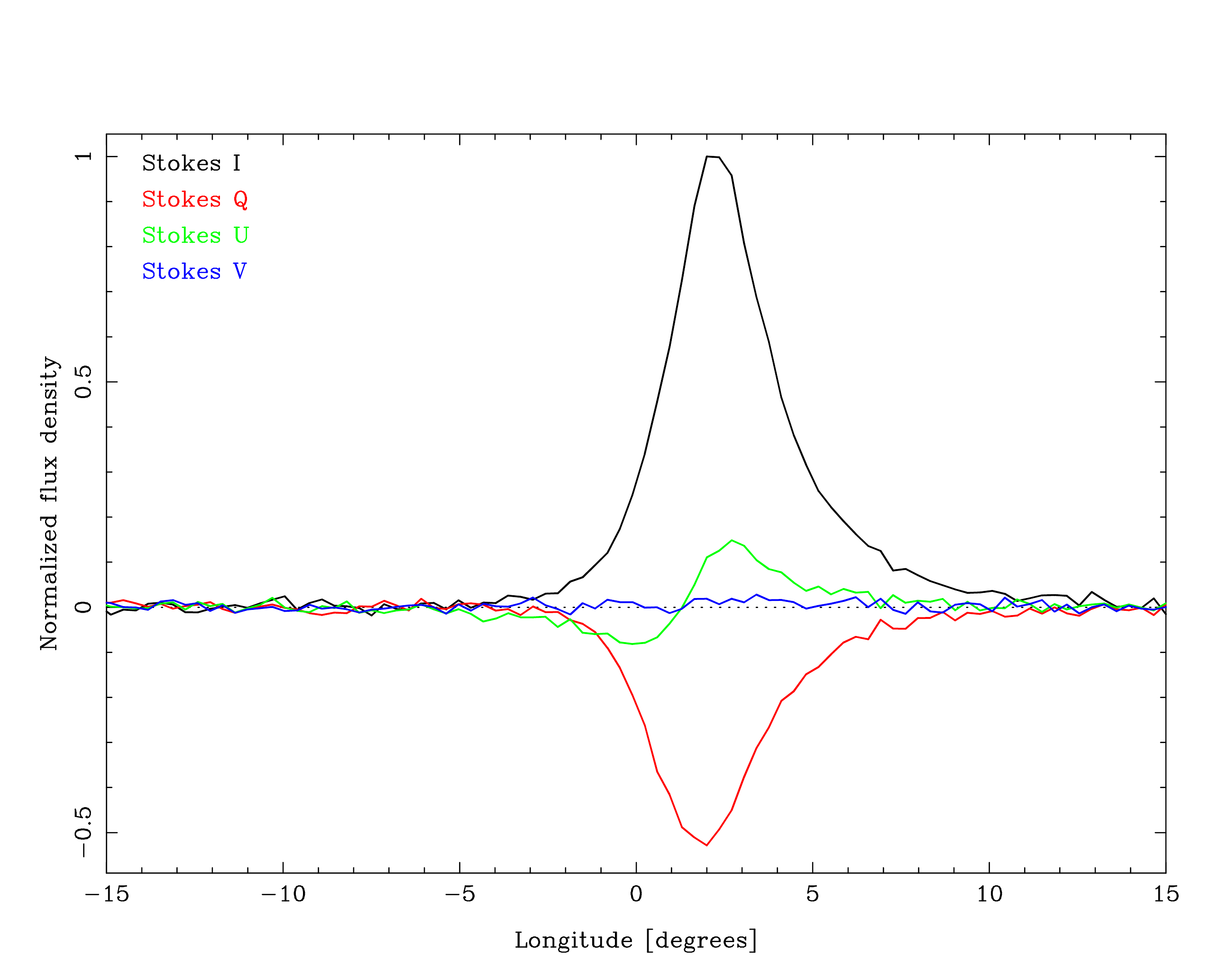}
\label{fig:subfig1}}%
\subfloat[Theoretical plots]{
\includegraphics[angle=0,width=8.45cm]{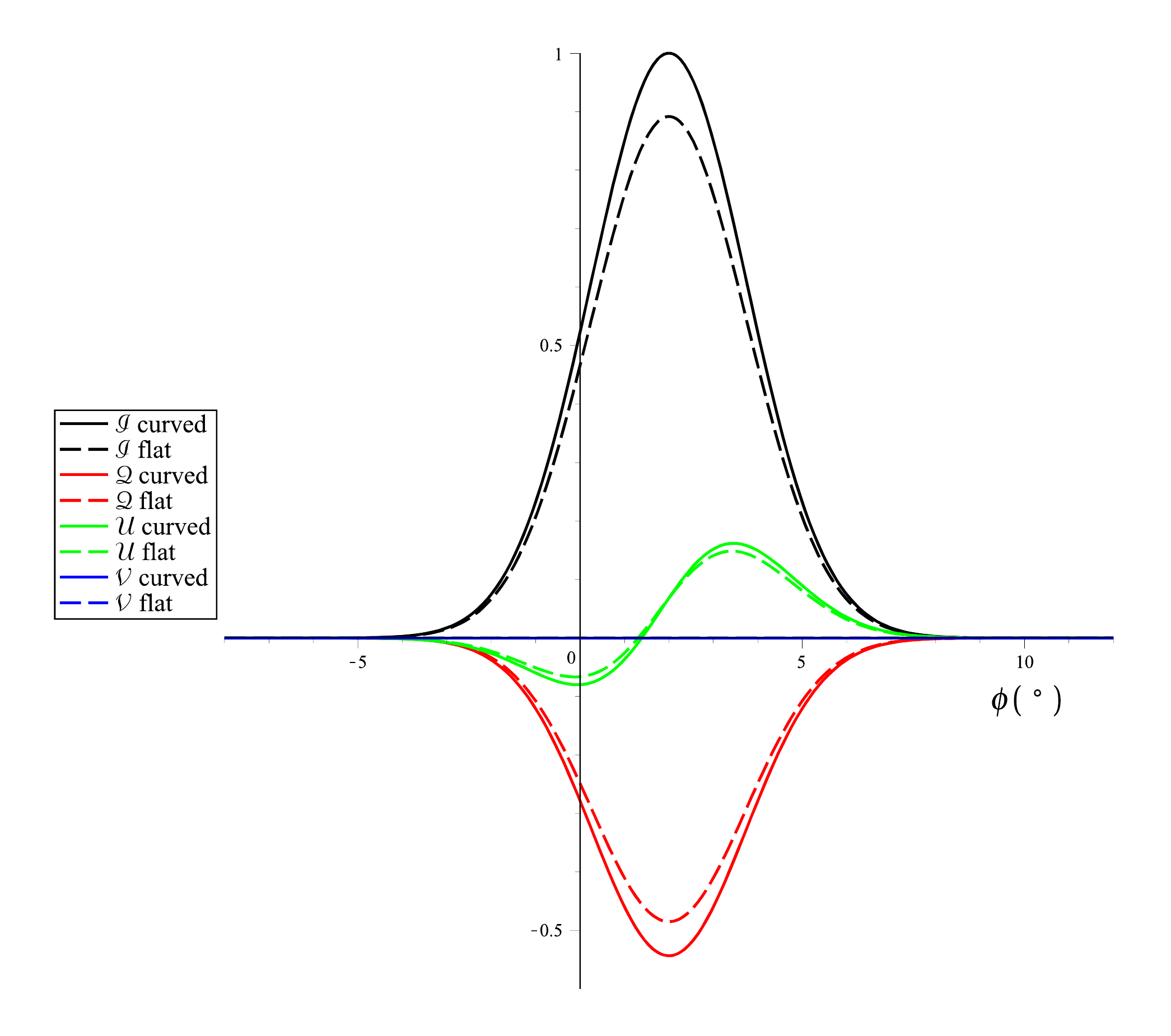}
\label{fig:subfig2}} 
\caption{Pulse profiles for the main pulse emission from PSR J1828-1101. }
\label{fig5}
\end{figure*}

\begin{figure*}[tbp]
\subfloat[Subfigure 1 list of figures text][Actual data plots]{
\includegraphics[angle=0,width=8.75cm]{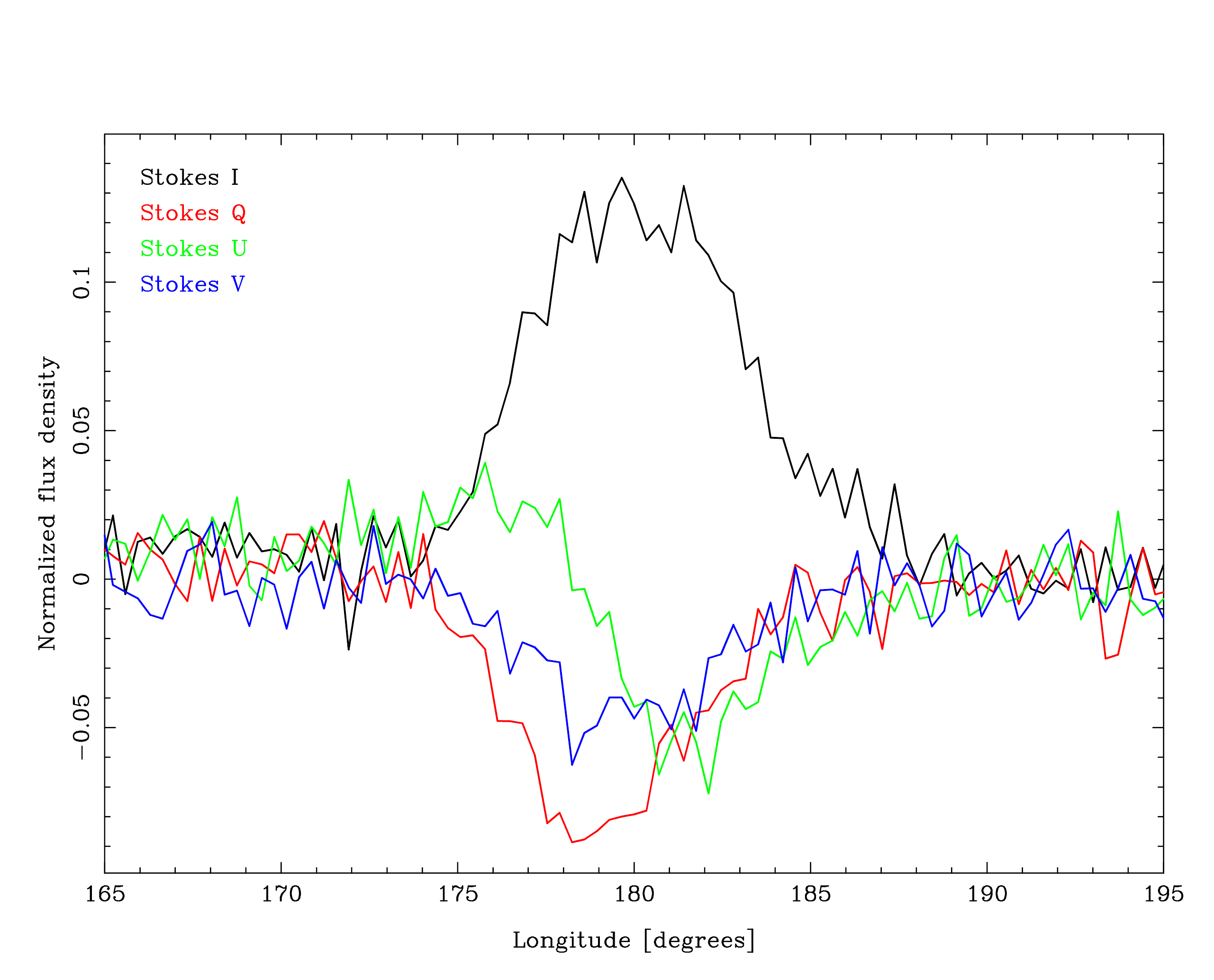}
\label{fig:subfig3}} 
\subfloat[Subfigure 2 list of figures text][Theoretical plots]{
\includegraphics[angle=0,width=9.1cm]{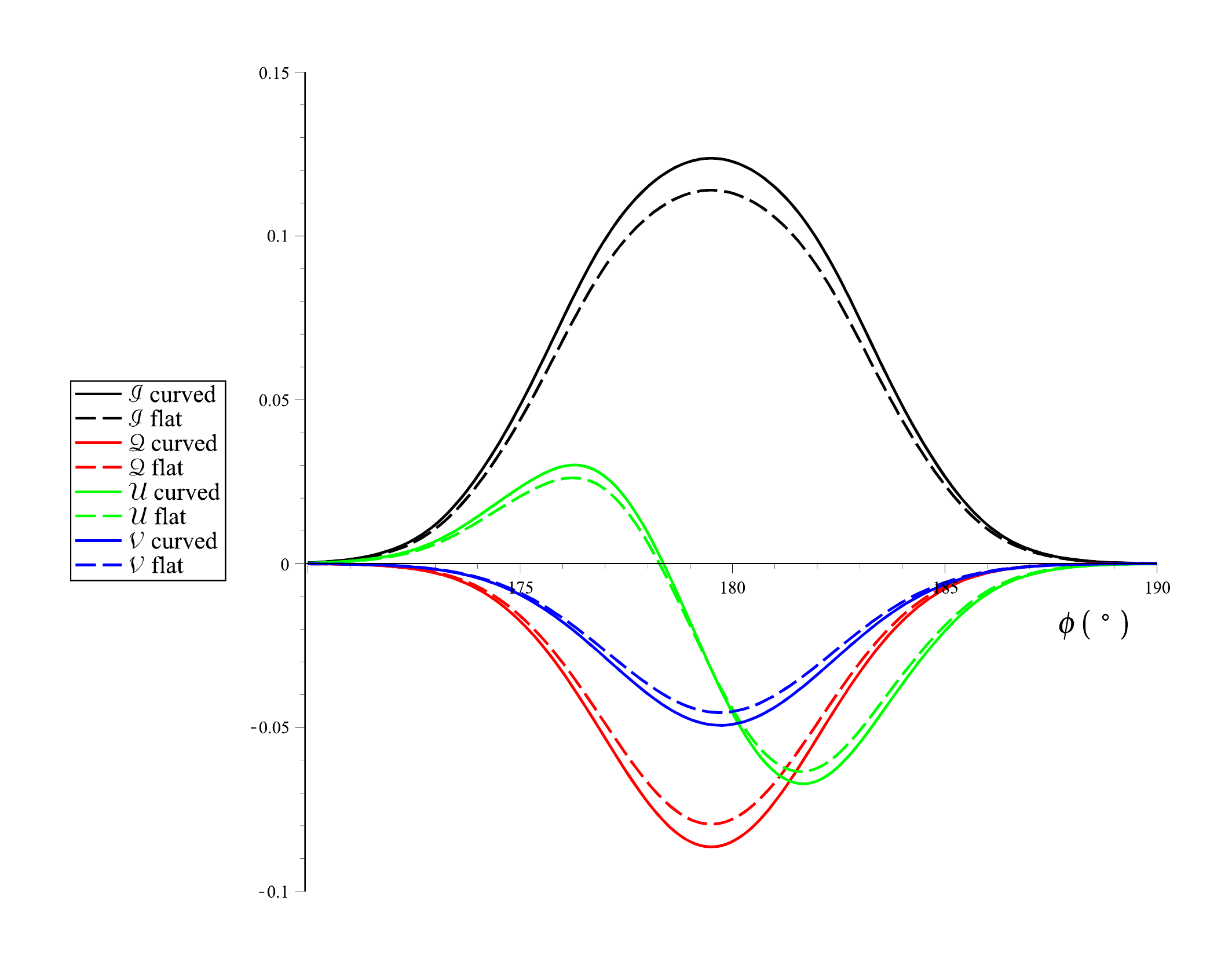}
\label{fig:subfig4}}
\caption{Pulse profiles for the interpulse\ emission from PSR J1828-1101. }
\label{fig6}
\end{figure*}

Further, the pulse profiles can be expressed as functions of the rotation
phase $\phi $ rather than the magnetic azimuthal phase $\varphi $. To this
end, one substitutes into Eqs. (\ref{pp1})-(\ref{pp4}) the conversion 
\begin{eqnarray}
\varphi &=&\arctan \left( \frac{\sin \theta \sin \phi }{\sin \alpha \cos
\theta -\cos \alpha \sin \theta \cos \phi }\right)  \notag \\
&\approx &\arctan \left( \frac{\sin \left( \alpha +\varepsilon \right) \sin
\phi }{\sin \alpha \cos \left( \alpha +\varepsilon \right) -\cos \alpha \sin
\left( \alpha +\varepsilon \right) \cos \phi }\right)  \notag \\
&&+\mathcal{O}\left( \left( \theta -\left( \alpha +\varepsilon \right)
\right) \right) ,  \label{pp8}
\end{eqnarray}%
where the approximation in the second line is based upon radiation emitted
from a single source charge at $\theta _{\mathrm{s}}=\alpha +\varepsilon $
on a magnetic field line, with $\alpha $ and $\varepsilon $ denoting the
inclination angle and the sight line impact angle \cite{Ganga2010},
respectively (see Fig. \ref{fig3} for illustration of $\alpha $ and $%
\varepsilon $).\footnote{%
Later, the pulse profile curves as shown in Figs. \ref{fig5} and \ref{fig6}
are plotted with an assumption of radiation emitted by a bunch of charges
rather than a single charge. Then, for a charge bunch centred at $\theta
=\alpha +\varepsilon $, the contribution from the term $\mathcal{O}\left(
\left( \theta -\left( \alpha +\varepsilon \right) \right) \right) $ would
not be negligible; in fact, the Gaussian modulation of this over $\theta $
will cause the pulse profile curves to shift upward or downward.}

The pulse profiles (\ref{pp1})-(\ref{pp4}) are dependent upon the curvature
radius $\rho $ through $\omega _{\mathrm{o}}$ given by Eq. (\ref{sp17}).
Hence, along with the conversion (\ref{pp8}) above, the curvature radius as
given by Eq. (\ref{sp14}) should also be rewritten such that its dependence
on the rotation phase $\phi $ is retrieved. It is expressed as 
\begin{widetext}
\begin{align}
\rho& =r_{\mathrm{s}}\left[ 1+\,\frac{g^{2}\left( r_{\mathrm{s}}\right) \left[
\cos \alpha \cos \theta +\sin \alpha \sin \theta \cos \phi \right] ^{2}}{1-\left[ \cos \alpha
\cos \theta +\sin \alpha \sin \theta \cos \phi \right] ^{2}}\right] ^{1/2}  \notag \\
& \approx r_{\mathrm{s}}\left[ 1+\,\frac{g^{2}\left( r_{\mathrm{s}}\right) \left[
\cos \alpha \cos \left( \alpha +\varepsilon \right) +\sin \alpha \sin \left(
\alpha +\varepsilon \right) \cos \phi \right] ^{2}}{1-\left[ \cos \alpha
\cos \left( \alpha +\varepsilon \right) +\sin \alpha \sin \left( \alpha
+\varepsilon \right) \cos \phi \right] ^{2}}\right] ^{1/2} +\mathcal{O}\left(\left(\theta -\left( \alpha +\varepsilon \right)\right)\right),  \label{pp9}
\end{align}
\end{widetext}where $g\left( r\right) $ refers to Eq. (\ref{dv9}): $%
g^{2}\left( r_{\mathrm{s}}\right) \leq 4$ with $g\left( r_{\mathrm{s}%
}\right) \rightarrow -2$ in the flat spacetime limit $m\rightarrow 0$. Here
again, the approximation in the second line is based upon radiation emitted
from a single source charge at $\theta _{\mathrm{s}}=\alpha +\varepsilon $
on a magnetic field line.

The technical details regarding the conversion (\ref{pp8}) and the curvature
radius (\ref{pp9}) are presented in Appendix \ref{appC}.

\begin{center}
\textbf{Example: PSR J1828-1101}
\end{center}

The pulsar PSR J1828-1101 is known to show `interpulse' emission, which is
nearly $180^{\circ }$ apart from its main pulse emission in rotation phase,
and hence is close to an orthogonal rotator with an inclination angle near $%
90^{\circ }$ \cite{JK2019}. Besides, this pulsar has a very important
property that we can exploit for our general relativistic analysis of pulsar
radio emission: its emission heights for both main and interpulse emissions
are fairly low; viz., a few times the radius of the neutron star, at which
gravity has a significant effect on the pulse profiles. In order to compare
theoretical and observed pulse profiles for this source, we obtained the
full-polarization pulse profiles observed by the Parkes radio telescope at $%
1.4\,\mathrm{GHz}$ and published by \cite{JK2019}.\footnote{%
Data were provided by S. Johnston in private communication.} The data are
summarized in Figs. \ref{fig5} and \ref{fig6}.

The pulse profiles for the emission from this pulsar can be plotted out of
the Stokes parameters $S=\left\{ I,Q,U,V\right\} $ given by Eqs. (\ref{pp1}%
)-(\ref{pp4}) against the rotation phase $\phi $ by means of Eqs. (\ref{sp17}%
), (\ref{pp8}) and (\ref{pp9}). These Stokes parameters are expressed using
the basis of linearly polarized waves \cite{Trippe2014}: 
\begin{equation}
\mathbf{e}_{\parallel }=\left[ 
\begin{array}{l}
1 \\ 
0%
\end{array}%
\right] ;~\mathbf{e}_{\perp }=\left[ 
\begin{array}{l}
0 \\ 
1%
\end{array}%
\right] ,  \label{ex1}
\end{equation}%
where $\mathbf{e}_{\parallel }$ and $\mathbf{e}_{\perp }$ denote linear
polarization in the directions parallel and perpendicular to the plane of
motion of the charge, respectively, as illustrated in Fig. \ref{fig4}.
However, the Stokes parameters can also be expressed using the basis $%
\left\{ \mathbf{e}_{1},\mathbf{e}_{2}\right\} =\left\{ \mathbf{e}_{\perp },%
\mathbf{e}_{\parallel }\right\} $, i.e., with the basis vectors swapped. In
this basis the Stokes parameters $S^{\prime }=\left\{ I^{\prime },Q^{\prime
},U^{\prime },V^{\prime }\right\} $ read%
\begin{align}
& I^{\prime }=\tilde{E}_{1}^{\ast }\tilde{E}_{1}+\tilde{E}_{2}^{\ast }\tilde{%
E}_{2}=\tilde{E}_{\perp }^{\ast }\tilde{E}_{\perp }+\tilde{E}_{\parallel
}^{\ast }\tilde{E}_{\parallel }=I,  \label{ex2} \\
& Q^{\prime }=\tilde{E}_{1}^{\ast }\tilde{E}_{1}-\tilde{E}_{2}^{\ast }\tilde{%
E}_{2}=\tilde{E}_{\perp }^{\ast }\tilde{E}_{\perp }-\tilde{E}_{\parallel
}^{\ast }\tilde{E}_{\parallel }=-Q,  \label{ex3} \\
& U^{\prime }=\tilde{E}_{1}^{\ast }\tilde{E}_{2}+\tilde{E}_{1}\tilde{E}%
_{2}^{\ast }=\tilde{E}_{\perp }^{\ast }\tilde{E}_{\parallel }+\tilde{E}%
_{\perp }\tilde{E}_{\parallel }^{\ast }=U,  \label{ex4} \\
& V^{\prime }=-\mathrm{i}\left( \tilde{E}_{1}^{\ast }\tilde{E}_{2}-\tilde{E}%
_{1}\tilde{E}_{2}^{\ast }\right) =-\mathrm{i}\left( \tilde{E}_{\perp }^{\ast
}\tilde{E}_{\parallel }-\tilde{E}_{\perp }\tilde{E}_{\parallel }^{\ast
}\right) =-V,  \label{ex5}
\end{align}%
where the third and the fourth equalities are consistent with the phase
difference, $\cos \left( \psi _{1}-\psi _{2}\right) =\cos \left( \psi
_{\perp }-\psi _{\parallel }\right) =\cos \psi $ and $\sin \left( \psi
_{1}-\psi _{2}\right) =\sin \left( \psi _{\perp }-\psi _{\parallel }\right)
=-\sin \psi $, in comparison with Eqs. (\ref{pp3}) and (\ref{pp4}),
respectively.

Further, one can consider the pulsar emission as coherent curvature
radiation by a bunch of charged particles, rather than single-particle
radiation. Taking this into consideration, our pulse profiles can be modeled
using a Gaussian modulation function \cite{GS1990,Ganga2010,KG2012}. One can
define new Stokes parameters $\mathcal{S}=\left\{ \mathcal{I},\mathcal{Q},%
\mathcal{U},\mathcal{V}\right\} $ as 
\begin{equation}
\mathcal{S}\left( \lambda \phi \right) \equiv \int S^{\prime }f\left( \theta
_{\star },\phi _{\star }\right) d\theta _{\star }d\phi _{\star },
\label{ex6}
\end{equation}%
where $\lambda $ is a free scaling factor to resize the rotation phase $\phi 
$\footnote{%
\red We absorbed two effects in the factor $\lambda $: (1) For a distant
observer of the radiation from a collimated bunch of mono-energetic charges
with Lorentz factor $\gamma $, the effective phase angle is $\gamma \sin
\phi $ (e.g., \cite{Trippe2014}), corresponding to $\gamma \phi $ in
small-angle approximation; (2) a distant observer receives emission from an
extended area on the surface of the neutron star, which leads to a spread of
the pulse over a range of phase angles.}, and $S^{\prime }=\left\{ I^{\prime
},Q^{\prime },U^{\prime },V^{\prime }\right\} $ refers to Eqs. (\ref{ex2})-(%
\ref{ex5}), and 
\begin{equation}
f\left( \theta _{\star },\phi _{\star }\right) =\frac{\exp \left( -\frac{%
\left( \theta _{\star }-\theta _{\mathrm{o}}\right) ^{2}}{\sigma _{\phi }^{2}%
}\right) \exp \left( -\frac{\left( \phi _{\star }-\phi _{\mathrm{o}}\right)
^{2}}{\sigma _{\phi }^{2}}\right) }{2\pi \sigma _{\theta }\sigma _{\phi
}\tau }  \label{ex7}
\end{equation}%
is the modulation function with $\sigma _{\theta }$ and $\sigma _{\phi }$
being the angular spread over $\theta $ and $\phi $, respectively, covering
a patch over a segment of a bundle of pulsar magnetic field lines, and $%
\left( \theta _{\mathrm{o}},\phi _{\mathrm{o}}\right) $ defining the peak
location of the function, equivalent to the centre of a charge bunch, and $%
\tau $ being a tuning factor to adjust the peak height. Now, with
consideration of (\ref{pp8}) and (\ref{pp9}) in regard to (\ref{ex2})-(\ref%
{ex5}), the expression (\ref{ex6}) can be approximated as\footnote{%
The approximation is employed to compute our theoretical pulse profiles in
Figs. \ref{fig5} and \ref{fig6}. However, while we compute Gaussian
modulations of $S^{\prime }\left( \phi -\phi _{\star }\right) $ explicitly,
only the effects of Gaussian modulations of $\mathcal{O}\left( \left( \theta
-\theta _{\star }\right) \right) $ are treated numerically, with no explicit
information about $\theta _{\mathrm{o}}$ and $\sigma _{\theta }$; we use
this effective method since the modulations of $\mathcal{O}\left( \left(
\theta -\theta _{\star }\right) \right) $ that depend on $\theta _{\mathrm{o}%
}$ and $\sigma _{\theta }$ are not well constrained by the observational
data.} 
\begin{equation}
\mathcal{S}\left( \lambda \phi \right) \approx \int \left[ S^{\prime }\left(
\phi -\phi _{\star }\right) +\mathcal{O}\left( \left( \theta -\theta _{\star
}\right) \right) \right] f\left( \theta _{\star },\phi _{\star }\right)
d\theta _{\star }d\phi _{\star }.  \label{ex8}
\end{equation}

In Fig. \ref{fig5} are presented the pulse profiles for the
\textquotedblleft main pulse\textquotedblright\ of curvature radiation from
PSR J1828-1101. Fig. \ref{fig:subfig1} shows the plots of $I$ (black), $Q$
(red), $U$ (green) and $V$ (blue) created based on the actual data obtained
from observations at $1.4\,\mathrm{GHz}$. Corresponding to this, Fig. \ref%
{fig:subfig2} shows our theoretical plots of $\mathcal{I}$ (black), $%
\mathcal{Q}$ (red), $\mathcal{U}$ (green) and $\mathcal{V}$ (blue) for the
emission by a Gaussian particle bunch in curved~(solid line) and flat
(dashed line) spacetimes: they are modeled with the emission height $r_{%
\mathrm{s}}\simeq 3.5\times 10^{6}\,\mathrm{cm}$; the inclination angle $%
\alpha =82^{\circ }.7$; the sight line impact angle $\varepsilon =7^{\circ
}.3$; the phase $\psi =0^{\circ }$; the angular spread $\sigma _{\phi
}=0^{\circ }.17$; the peak location $\phi _{\mathrm{o}}=2^{\circ }$; the
tuning factor $\tau =1$; the scaling factor $\lambda =\frac{1}{10}$ {%
\red
(which matches the model pulse width to the observed width)}, for the
neutron star mass $M\simeq 1.4\,M_{\odot }$ ($m=GM/c^{2}\simeq 2.065\times
10^{5}\,\mathrm{cm}$); the Lorentz factor $\gamma =400$; the observation
frequency $\omega =1.4\times 10^{9}\,\mathrm{Hz}$.\footnote{%
In Figs. \ref{fig5} and \ref{fig6} our pulse profile curves are plotted with
a vertical axis scale relative to the maximum of the intensity $\mathcal{I}$
for the main pulse in curved spacetime, which is set to be $1$.}\footnote{%
The Gaussian modulations of $\mathcal{O}\left( \left( \theta -\theta _{\star
}\right) \right) $ from (\ref{ex8})\ have been treated numerically. Among
others, the effect of the modulation for the Stokes $U$ is notable; it has
caused the pulse profile curve to shift upward \cite{KG2012}, which has
moved the zero of the curve to the left as a result.}

Similarly, in Fig. \ref{fig6} are presented the pulse profiles for the
\textquotedblleft interpulse\textquotedblright\ of curvature radiation from
PSR J1828-1101. Fig. \ref{fig:subfig3} shows the plots of $I$ (black), $Q$
(red), $U$ (green) and $V$ (blue) created based on the actual data obtained
from observations at $1.4\,\mathrm{GHz}$. Corresponding to this, Fig. \ref%
{fig:subfig4} shows our theoretical plots of $\mathcal{I}$ (black), $%
\mathcal{Q}$ (red), $\mathcal{U}$ (green) and $\mathcal{V}$ (blue) for the
emission by a Gaussian particle bunch in curved~(solid line) and flat
(dashed line) spacetimes: they are modeled with the emission height $r_{%
\mathrm{s}}\simeq 4.5\times 10^{6}\,\mathrm{cm}$; the inclination angle $%
\alpha =97^{\circ }.3$; the sight line impact angle $\varepsilon =-7^{\circ
}.3$; the phase $\psi =45^{\circ }$; the angular spread $\sigma _{\phi
}=0^{\circ }.03$; the peak location $\phi _{\mathrm{o}}=179^{\circ }.5$; the
tuning factor $\tau =10$; the scaling factor $\lambda =\frac{1}{60}$ {%
\red
(which matches the model pulse width to the observed width)}, for the
same $M $, $\gamma $ and $\omega $.\footnote{%
Similarly, the Gaussian modulations of $\mathcal{O}\left( \left( \theta
-\theta _{\star }\right) \right) $ from (\ref{ex8})\ have been treated
numerically. Again, the effect of the modulation for the Stokes $U$ is
notable; it has caused the pulse profile curve to shift downward \cite%
{KG2012}, which has moved the zero of the curve to the left as a result.} 

Note that for Figs. \ref{fig:subfig2} and \ref{fig:subfig4} the values of $%
r_{\mathrm{s}}$, $\alpha $, $\varepsilon $ and $\omega $ have been taken
from Ref. \cite{JK2019}, whereas the values of $\psi $, $\sigma _{\phi }$, $%
\phi _{\mathrm{o}}$, $\tau $, $\lambda $ and $\gamma $ have been chosen such
that our theoretical plots match best with observational plots in Figs. \ref%
{fig:subfig1} and \ref{fig:subfig3}. Also, it should be noted that circular
polarization as exhibited by $V$ (blue) in Fig. \ref{fig:subfig3} is
entirely negative; hence, our $\mathcal{V}$ (blue) in Fig. \ref{fig:subfig4}
has been modeled following this. Here our method can be described as
follows. From Eq. (\ref{ex5}), one can reexpress $V^{\prime }=-V=E_{\mathrm{L%
}}^{\ast }E_{\mathrm{L}}-E_{\mathrm{R}}^{\ast }E_{\mathrm{R}}$ in the basis $%
\left\{ \mathbf{e}_{\mathrm{L}},\mathbf{e}_{\mathrm{R}}\right\} $, where $%
\mathbf{e}_{\mathrm{L}}=\left[ 
\begin{array}{ll}
1 & \mathrm{i}%
\end{array}%
~\right] ^{\mathrm{T}}$ and $\mathbf{e}_{\mathrm{R}}=\left[ 
\begin{array}{ll}
1 & -\mathrm{i}%
\end{array}%
\right] ^{\mathrm{T}}$ denote left-hand and right-hand circular
polarization, respectively \cite{Trippe2014}. However, in order to have the
entirely negative support over the whole domain of $\phi $, $V^{\prime }$
has to be reduced to $-pE_{\mathrm{R}}^{\ast }E_{\mathrm{R}}=-\frac{p}{2}%
I^{\prime }+\frac{p}{2}V^{\prime }$, where $0<p\leq 1$; i.e., it has to be
right-hand circularly polarized only.\footnote{%
Suppose that $V^{\prime }$ is reduced to $v\equiv p_{1}\tilde{E}_{\mathrm{L}%
}^{\ast }\tilde{E}_{\mathrm{L}}-p_{2}\tilde{E}_{\mathrm{R}}^{\ast }\tilde{E}%
_{\mathrm{R}}$, where $0\leq p_{1},p_{2}\leq 1$. Now, with $\tilde{E}_{%
\mathrm{L}}=\frac{1}{\sqrt{2}}\left( \tilde{E}_{\parallel }+\mathrm{i}\tilde{%
E}_{\perp }\right) $ and $\tilde{E}_{\mathrm{R}}=\frac{1}{\sqrt{2}}\left( 
\tilde{E}_{\parallel }-\mathrm{i}\tilde{E}_{\perp }\right) $, we can express 
$v=\frac{p_{1}-p_{2}}{2}I^{\prime }+\frac{p_{1}+p_{2}}{2}V^{\prime }$, due
to Eqs. (\ref{ex2}) and (\ref{ex5}). However, if $v\leq 0$ everywhere in the
domain of $\phi $, the following must hold true always: $\frac{p_{1}-p_{2}}{%
p_{1}+p_{2}}\left( X^{2}+Y^{2}\right) -2\sin \psi XY\leq 0$, where $%
p_{1}<p_{2}$, and $X=\mathcal{E}_{\mathrm{o}}\omega \varphi \left( \delta
^{2}+\varphi ^{2}\right) \mathrm{K}_{1/3}\left( \frac{\omega }{3\omega _{%
\mathrm{o}}}\left( \delta ^{2}+\varphi ^{2}\right) ^{3/2}\right) $ and $Y=%
\mathcal{E}_{\mathrm{o}}\omega \left( \delta ^{2}+\varphi ^{2}\right) 
\mathrm{K}_{2/3}\left( \frac{\omega }{3\omega _{\mathrm{o}}}\left( \delta
^{2}+\varphi ^{2}\right) ^{3/2}\right) $, defined via Eqs. (\ref{pp1}), (\ref%
{pp4}), (\ref{ex2}) and (\ref{ex5}). This inequality can be satisfied only
if $p_{1}=0$ and $p_{2}>0$; i.e., $-\sin \psi \left( X+Y\right) ^{2}\leq
\left( 1-\sin \psi \right) \left( X^{2}+Y^{2}\right) $ for $0^{\circ }\leq
\psi \leq 90^{\circ }$. Therefore, $V^{\prime }$ is eventually reduced to $%
v=-p_{2}\tilde{E}_{\mathrm{R}}^{\ast }\tilde{E}_{\mathrm{R}}$, where $%
0<p_{2}\leq 1$.} To have the best match with $V$ (blue) in Fig. \ref%
{fig:subfig3}, we have chosen $p\approx 0.8$ and thus plotted our $\mathcal{V%
}$ (blue) by modulating \thinspace $-0.4I^{\prime }+0.4V^{\prime }$ with the
Gaussian function via Eq. (\ref{ex6}).

\section{Summary and Conclusions}

We have investigated general relativistic effects in electromagnetism of a
pulsar based on its magnetic dipole model defined in curved spacetime. Our
analysis can be briefly summarized as follows. The magnetic field is
described by a solution to Maxwell's equations in a spacetime geometry
specially prescribed for an oblique rotator: a slowly rotating Schwarzschild
geometry modified from the Kerr spacetime, with the frame-dragging frequency
replaced by the pulsar rotation frequency (Sections \ref{trans}-\ref%
{reference}). The magnetic field in the curved spacetime manifestly exhibits
the effect of gravity: the stronger the closer to the surface of the neutron
star. By means of the magnetic field in the curved spacetime, well-known
issues such as very low-frequency magnetic dipole radiation for pulsar
spin-down (Section \ref{radiative}) and pulse profiles of curvature
radiation (Section \ref{profile}) are extended to the context of general
relativity. Also, these results are compared with their flat-spacetime
counterparts so that the differences manifestly show the effects of gravity.

In this study, we have devoted considerable attention to the mechanism of
pulsar radio emission in the context of general relativity: among other
things, our primary focus has been on general relativistic effects on the
pulse profiles of curvature emission. Well inside a pulsar magnetosphere,
the magnetic field strength is very high, and so is the strength of gravity
of the neutron star. The effect of gravity is so intense as to modify the
magnetic field lines in this region: the curvature radii of the field lines
would decrease due to strong gravity (Eqs. (\ref{sp13}) and (\ref{sp14})).
This would increase the frequency of curvature radiation emitted by charges
moving along the magnetic field lines in the region (Eq. (\ref{sp17})), and
in turn, modify the pulse profiles (Eqs. (\ref{pp1})-(\ref{pp4})). Figs. \ref%
{fig5} and \ref{fig6} show this clearly. Overall, the effect of gravity
increases the magnitude of our pulse profile curves for intensity $\mathcal{I%
}$, linear polarization $\mathcal{Q}$ and $\mathcal{U}$ , and circular
polarization $\mathcal{V}$ modeled for Gaussian particle-bunch radiation:
from Eqs. (\ref{sp17}) and (\ref{pp1})-(\ref{pp4}) $\mathrm{magnitude}\sim 
\mathcal{E}_{\mathrm{o}}^{2}\sim \omega _{\mathrm{o}}^{2}\sim \rho ^{-2}$,
and therefore $\mathrm{magnitude~(curved)}>\mathrm{magnitude~(flat)}$. We
have taken the pulsar PSR J1828-1101 as an example, in which the pulse
profiles of the main and interpulse emissions from the regions $3.5R_{\ast
}\simeq $ $3.5\times 10^{6}\,\mathrm{cm}$ and $4.5R_{\ast }\simeq $ $%
4.5\times 10^{6}\,\mathrm{cm} $, respectively, above the surface of the
neutron star of mass $M\simeq 1.4\,M_{\odot }$ ($m=GM/c^{2}\simeq
2.065\times 10^{5}\,\mathrm{cm}$) have shown that $\mathrm{magnitude~(flat)}$
is about $89\,\%$ and $92\,\%$ of $\mathrm{magnitude~(curved)}$ for the main
pulse and the interpulse, respectively. This clearly exhibits the effect of
gravity, which is due to the locations of emissions being relatively close
to the surface of the star, where gravity is strong enough to affect
electromagnetism of the pulsar.

Through our analysis based on general relativity, it is of great interest to
see the gravitational effects in electromagnetism of a pulsar; especially,
the effects on the pulse profiles. However, while actual observations of
pulsars are indeed believed to contain these effects, it would be regarded
as extremely difficult to disentangle and to identify them alone from
observational data. Despite theoretical interest, the effects make a
difference by $8\,\%$ to $11\,\%$ only, and therefore their testability is
beyond the current detection capabilities since the absolute magnitude of
the pulse profiles is not precisely predictable; not even the order of
magnitude. We leave a discussion of this issue to a follow-up study.

\section*{Acknowledgments}

D.-H. Kim acknowledges financial support from the National Research
Foundation of Korea (NRF) via Basic Research Grants NRF-2018R1D1A1B07051276
and NRF-2021R1I1A1A01054781. S. Trippe acknowledges financial support from
the NRF via Basic Research Grant NRF-2019R1F1A1059721. We are grateful to
Dr. Simon Johnston for providing the data for J1828-1101. The Parkes
telescope is part of the Australia Telescope National Facility which is
funded by the Commonwealth of Australia for operation as a National Facility
managed by CSIRO.

\appendix

\section{Transformation between the Inertial Frame and the Corotating Frame}

\label{appA}

In the Cartesian representation, we have the relation between the two
frames, $\mathbf{x}\equiv \left( x,y,z\right) $ (inertial frame) and $%
\mathbf{x}^{\prime }\equiv \left( x^{\prime },y^{\prime },z^{\prime }\right) 
$ (obliquely corotating frame): 
\begin{align}
& x^{\prime }=-\sin \left( \Omega t\right) x+\cos \left( \Omega t\right) y,
\label{x1} \\
& y^{\prime }=-\cos \alpha \cos \left( \Omega t\right) x-\cos \alpha \sin
\left( \Omega t\right) y+\sin \alpha \,z,  \label{y1} \\
& z^{\prime }=\sin \alpha \cos \left( \Omega t\right) x+\sin \alpha \sin
\left( \Omega t\right) y+\cos \alpha \,z.  \label{z1}
\end{align}%
Fig. \ref{fig2} illustrates how these two frames are related to each other.

By means of Eq. (\ref{z1}) and 
\begin{equation}
\cos \theta ^{\prime }=\frac{z^{\prime }}{r}  \label{tr1}
\end{equation}%
together with 
\begin{eqnarray}
x &=&r\sin \theta \cos \phi ,  \label{tr2} \\
y &=&r\sin \theta \sin \phi ,  \label{tr3} \\
z &=&r\cos \theta ,  \label{tr4}
\end{eqnarray}%
one can establish%
\begin{equation}
\cos \theta ^{\prime }=\cos \alpha \cos \theta +\sin \alpha \sin \theta \cos
\left( \phi -\Omega t\right) .  \label{tr5}
\end{equation}%
This is known as the spherical law of cosines, which can alternatively be
obtained from consideration of a spherical triangle on the surface of the
2-sphere. In Section \ref{oblique}, it gives a definition of the
\textquotedblleft magnetic colatitude\textquotedblright .

Equivalent to Eqs. (\ref{x1})-(\ref{z1}) are the relations between the basis
vectors:%
\begin{equation}
\left[ 
\begin{array}{c}
\mathbf{e}_{x^{\prime }} \\ 
\mathbf{e}_{y^{\prime }} \\ 
\mathbf{e}_{z^{\prime }}%
\end{array}%
\right] =\mathbf{T}\left[ 
\begin{array}{c}
\mathbf{e}_{x} \\ 
\mathbf{e}_{y} \\ 
\mathbf{e}_{z}%
\end{array}%
\right] ,  \label{e1}
\end{equation}%
where 
\begin{equation}
\mathbf{T}\equiv \left[ 
\begin{array}{ccc}
-\sin \left( \Omega t\right) & \cos \left( \Omega t\right) & 0 \\ 
-\cos \alpha \cos \left( \Omega t\right) & -\cos \alpha \sin \left( \Omega
t\right) & \sin \alpha \\ 
\sin \alpha \cos \left( \Omega t\right) & \sin \alpha \sin \left( \Omega
t\right) & \cos \alpha%
\end{array}%
\right] .  \label{T}
\end{equation}

A set of basis vectors in spherical polar coordinates $\left( \mathbf{e}_{%
\hat{r}},\mathbf{e}_{\hat{\theta}},\mathbf{e}_{\hat{\phi}}\right) $ can be
obtained by rotating the Cartesian set $\left( \mathbf{e}_{x},\mathbf{e}_{y},%
\mathbf{e}_{z}\right) $:%
\begin{equation}
\left[ 
\begin{array}{c}
\mathbf{e}_{\hat{r}} \\ 
\mathbf{e}_{\hat{\theta}} \\ 
\mathbf{e}_{\hat{\phi}}%
\end{array}%
\right] =\mathbf{R}\left[ 
\begin{array}{c}
\mathbf{e}_{x} \\ 
\mathbf{e}_{y} \\ 
\mathbf{e}_{z}%
\end{array}%
\right] ,  \label{e2}
\end{equation}%
where 
\begin{equation}
\mathbf{R}=\left[ 
\begin{array}{ccc}
\sin \theta \cos \phi & \sin \theta \sin \phi & \cos \theta \\ 
\cos \theta \cos \phi & \cos \theta \sin \phi & -\sin \theta \\ 
-\sin \phi & \cos \phi & 0%
\end{array}%
\right] .  \label{R}
\end{equation}%
The same argument holds true between $\left( \mathbf{e}_{\hat{r}},\mathbf{e}%
_{\hat{\theta}^{\prime }},\mathbf{e}_{\hat{\phi}^{\prime }}\right) $ and $%
\left( \mathbf{e}_{x^{\prime }},\mathbf{e}_{y^{\prime }},\mathbf{e}%
_{z^{\prime }}\right) $, with $\left( \theta ,\phi \right) $ replaced by $%
\left( \theta ^{\prime },\phi ^{\prime }\right) $: 
\begin{equation}
\left[ 
\begin{array}{c}
\mathbf{e}_{\hat{r}} \\ 
\mathbf{e}_{\hat{\theta}^{\prime }} \\ 
\mathbf{e}_{\hat{\phi}^{\prime }}%
\end{array}%
\right] =\mathbf{R}^{\prime }\left[ 
\begin{array}{c}
\mathbf{e}_{x^{\prime }} \\ 
\mathbf{e}_{y^{\prime }} \\ 
\mathbf{e}_{z^{\prime }}%
\end{array}%
\right] ,  \label{e3}
\end{equation}%
where 
\begin{equation}
\mathbf{R}^{\prime }=\left[ 
\begin{array}{ccc}
\sin \theta ^{\prime }\cos \phi ^{\prime } & \sin \theta ^{\prime }\sin \phi
^{\prime } & \cos \theta ^{\prime } \\ 
\cos \theta ^{\prime }\cos \phi ^{\prime } & \cos \theta ^{\prime }\sin \phi
^{\prime } & -\sin \theta ^{\prime } \\ 
-\sin \phi ^{\prime } & \cos \phi ^{\prime } & 0%
\end{array}%
\right] .  \label{R2}
\end{equation}

Combining the transformations given by Eqs. (\ref{T}), (\ref{R}) and (\ref%
{R2}), and after somewhat complicated algebraic manipulation, the
transformation between the coordinate frames $\left( r,\theta ,\phi \right) $
and $\left( r,\theta ^{\prime },\phi ^{\prime }\right) $ can be finally
determined:%
\begin{equation}
\left[ 
\begin{array}{c}
\mathbf{e}_{\hat{r}} \\ 
\mathbf{e}_{\hat{\theta}} \\ 
\mathbf{e}_{\hat{\phi}}%
\end{array}%
\right] =\mathbf{M}\left[ 
\begin{array}{c}
\mathbf{e}_{\hat{r}} \\ 
\mathbf{e}_{\hat{\theta}^{\prime }} \\ 
\mathbf{e}_{\hat{\phi}^{\prime }}%
\end{array}%
\right] ,  \label{e4}
\end{equation}%
where%
\begin{widetext}
\begin{eqnarray}
\mathbf{M} &=&\mathbf{RT}^{-1}\mathbf{R}^{\prime -1}  \notag \\
&=&\left[ 
\begin{array}{ccc}
1 & 0 & 0 \\ 
0 & \frac{\cos \alpha \sin \theta -\sin \alpha \cos \theta \cos \left( \phi
-\Omega t\right) }{\sin \theta ^{\prime }} & \frac{-\sin \alpha \sin \left(
\phi -\Omega t\right) }{\sin \theta ^{\prime }} \\ 
0 & \frac{\sin \alpha \sin \left( \phi -\Omega t\right) }{\sin \theta
^{\prime }} & \frac{\cos \alpha \sin \theta -\sin \alpha \cos \theta \cos
\left( \phi -\Omega t\right) }{\sin \theta ^{\prime }}%
\end{array}%
\right] ,  \label{MA}
\end{eqnarray}
\end{widetext}
and $\theta ^{\prime }$ is defined through Eq. (\ref{tr5}).

\section{Description of Motion of a Charge in a Specially Chosen Cartesian
Frame}

\label{appB}

In Section \ref{curv} the trajectory $\mathbf{\xi }_{\mathrm{s}}$, the
velocity $\mathbf{\dot{\xi}}_{\mathrm{s}}$ and the acceleration $\mathbf{%
\ddot{\xi}}_{\mathrm{s}}$ of a source charge along a magnetic field line are
reexpressed in a Cartesian frame specially chosen for computational
convenience, as given by Eqs. (\ref{sp20-1})-(\ref{sp22}). These Cartesian
expressions can be obtained through multiple coordinate transformations of
the initial spherical polar representations (\ref{sp5})-(\ref{sp7}).

The spherical polar representations of $\mathbf{\xi }_{\mathrm{s}}$, $%
\mathbf{\dot{\xi}}_{\mathrm{s}}$ and $\mathbf{\ddot{\xi}}_{\mathrm{s}}$ in
Eqs. (\ref{sp5})-(\ref{sp7}) can be projected into a Cartesian frame via Eq.
(\ref{e3}) with $\phi ^{\prime }=0$:%
\begin{equation}
\mathbf{\xi }_{\mathrm{s}}=r_{\mathrm{s}}\left( \sin \theta _{\mathrm{s}%
}^{\prime },0,\cos \theta _{\mathrm{s}}^{\prime }\right) ,  \label{xi1}
\end{equation}%
\begin{align}
& \mathbf{\dot{\xi}}_{\mathrm{s}}=\beta c  \notag \\
& \times \left( \frac{\left[ 1-g\left( r_{\mathrm{s}}\right) \right] \sin
\theta _{\mathrm{s}}^{\prime }}{\sqrt{g^{2}\left( r_{\mathrm{s}}\right)
+\tan ^{2}\theta _{\mathrm{s}}^{\prime }}},0,\frac{-\left[ g\left( r_{%
\mathrm{s}}\right) +\tan ^{2}\theta _{\mathrm{s}}^{\prime }\right] \cos
\theta _{\mathrm{s}}^{\prime }}{\sqrt{g^{2}\left( r_{\mathrm{s}}\right)
+\tan ^{2}\theta _{\mathrm{s}}^{\prime }}}\right) ,  \label{xi2}
\end{align}%
\begin{align}
& \mathbf{\ddot{\xi}}_{\mathrm{s}}=\frac{\beta ^{2}c^{2}}{\rho }  \notag \\
& \times \left( \frac{-\left[ g\left( r_{\mathrm{s}}\right) +\tan ^{2}\theta
_{\mathrm{s}}^{\prime }\right] \cos \theta _{\mathrm{s}}^{\prime }}{\sqrt{%
g^{2}\left( r_{\mathrm{s}}\right) +\tan ^{2}\theta _{\mathrm{s}}^{\prime }}}%
,0,\frac{-\left[ 1-g\left( r_{\mathrm{s}}\right) \right] \sin \theta _{%
\mathrm{s}}^{\prime }}{\sqrt{g^{2}\left( r_{\mathrm{s}}\right) +\tan
^{2}\theta _{\mathrm{s}}^{\prime }}}\right) .  \label{xi3}
\end{align}%
From this we find $\mathbf{\dot{\xi}}_{\mathrm{s}}\cdot \mathbf{\ddot{\xi}}_{%
\mathrm{s}}=0$ but $\mathbf{\xi }_{\mathrm{s}}\cdot \mathbf{\dot{\xi}}_{%
\mathrm{s}}\neq 0$; that is, $\mathbf{\dot{\xi}}_{\mathrm{s}}$ and $\mathbf{%
\ddot{\xi}}_{\mathrm{s}}$ are perpendicular to each other, but $\mathbf{\xi }%
_{\mathrm{s}}$ and $\mathbf{\dot{\xi}}_{\mathrm{s}}$\ are not, which results
in $-\mathbf{\xi }_{\mathrm{s}}$ and $\mathbf{\ddot{\xi}}_{\mathrm{s}}$ not
being parallel to each other.

Now, in order to redefine $\mathbf{\xi }_{\mathrm{s}}$, $\mathbf{\dot{\xi}}_{%
\mathrm{s}}$ and $\mathbf{\ddot{\xi}}_{\mathrm{s}}$ in another Cartesian
frame, one can transform%
\begin{equation}
\mathbf{\xi }_{\mathrm{s}}^{\mathrm{N}}=\mathbf{\xi }_{\mathrm{s}}-%
\overrightarrow{OO^{\mathrm{N}}},  \label{xi4}
\end{equation}%
where $O\equiv \left( 0,0,0\right) $ is the fixed origin, and $O^{\mathrm{N}%
} $ is a new origin, which makes the centre for an instantaneous circle of
radius $\rho $ passing through the point $\mathbf{\xi }_{\mathrm{s}}$ given
by (\ref{xi1}) (see Fig. \ref{fig4}). Through some analysis, it can be shown
that 
\begin{equation}
\overrightarrow{OO^{\mathrm{N}}}=\left( r_{\mathrm{s}}\sin \theta _{\mathrm{s%
}}^{\prime }+\,\rho \cos \theta _{\mathrm{s}}^{\prime },0,r_{\mathrm{s}}\cos
\theta _{\mathrm{s}}^{\prime }-\,\rho \sin \theta _{\mathrm{s}}^{\prime
}\right) ,  \label{xi5}
\end{equation}%
which is not fixed, but varies with $t_{\mathrm{R}}$ through $r_{\mathrm{s}%
}\left( t_{\mathrm{R}}\right) $ and\ $\theta _{\mathrm{s}}^{\prime }\left(
t_{\mathrm{R}}\right) $. Then by Eqs. (\ref{xi1}), (\ref{xi4}) and (\ref{xi5}%
) we have 
\begin{equation}
\mathbf{\xi }_{\mathrm{s}}^{\mathrm{N}}=\left( -\rho \cos \theta _{\mathrm{s}%
}^{\prime },0,\rho \sin \theta _{\mathrm{s}}^{\prime }\right) .  \label{xi6}
\end{equation}%
With the substitution $\theta _{\mathrm{s}}^{\prime }=\vartheta +\pi /2$, we
finally obtain%
\begin{equation}
\mathbf{\xi }_{\mathrm{s}}^{\mathrm{N}}=\rho \left( \sin \vartheta ,0,\cos
\vartheta \right) .  \label{xi7}
\end{equation}%
Differentiating both sides of Eq. (\ref{xi7}) with respect to $t_{\mathrm{R}%
} $, we have%
\begin{equation}
\mathbf{\dot{\xi}}_{\mathrm{s}}^{\mathrm{N}}=\rho \left( \dot{\vartheta}\cos
\vartheta ,0,-\dot{\vartheta}\sin \vartheta \right) .  \label{xi8}
\end{equation}%
With $\theta _{\mathrm{s}}^{\prime }\left( t_{\mathrm{R}}\right) =\vartheta
\left( t_{\mathrm{R}}\right) +\pi /2$, Eq. (\ref{sp17}) leads to 
\begin{equation}
\dot{\vartheta}=\frac{\beta c}{\rho }.  \label{xi9}
\end{equation}%
Then using this for Eq. (\ref{xi8}), we obtain%
\begin{equation}
\mathbf{\dot{\xi}}_{\mathrm{s}}^{\mathrm{N}}=\beta c\left( \cos \vartheta
,0,-\sin \vartheta \right) .  \label{xi10}
\end{equation}%
Similarly, we further obtain%
\begin{equation}
\mathbf{\ddot{\xi}}_{\mathrm{s}}^{\mathrm{N}}=\beta c\left( -\dot{\vartheta}%
\sin \vartheta ,0,-\dot{\vartheta}\cos \vartheta \right) =-\frac{\beta
^{2}c^{2}}{\rho }\left( \sin \vartheta ,0,\cos \vartheta \right) .
\label{xi11}
\end{equation}

Dropping the superscript $^{\mathrm{N}}$ from Eqs. (\ref{xi7}), (\ref{xi10})
and (\ref{xi11}), we have exactly reproduced the expressions (\ref{sp20-1})-(%
\ref{sp22}). Note that in this representation $\mathbf{\xi }_{\mathrm{s}%
}\cdot \mathbf{\dot{\xi}}_{\mathrm{s}}=0$ and $\mathbf{\dot{\xi}}_{\mathrm{s}%
}\cdot \mathbf{\ddot{\xi}}_{\mathrm{s}}=0$; that is, $\mathbf{\xi }_{\mathrm{%
s}}\perp \mathbf{\dot{\xi}}_{\mathrm{s}}$, $\mathbf{\dot{\xi}}_{\mathrm{s}%
}\perp \mathbf{\ddot{\xi}}_{\mathrm{s}}$, and $-\mathbf{\xi }_{\mathrm{s}%
}\parallel \mathbf{\ddot{\xi}}_{\mathrm{s}}$. This results from redefining $%
\mathbf{\xi }_{\mathrm{s}}$, $\mathbf{\dot{\xi}}_{\mathrm{s}}$ and $\mathbf{%
\ddot{\xi}}_{\mathrm{s}}$ in the special Cartesian frame by moving the
origin from $O$\ to $O^{\mathrm{N}}$, following Eq. (\ref{xi4}).

\section{Pulse Profiles Expressed in the Rotation Phase $\protect\phi $}

\label{appC}

In Section \ref{curv} $\mathbf{\hat{r}}$, the unit vector for the
observational direction for a distant observer is given by Eq. (\ref{sp23}),
being parametrized by the azimuthal angle $\varphi $, as expressed in the
special Cartesian frame introduced in Appendix \ref{appB}. This is in
contrast to the motion of a source charge along a magnetic field line, which
is parametrized by the polar angle $\vartheta $. However, in consideration
of the analyses in Appendices \ref{appA} and \ref{appB}, one can identify $%
\theta ^{\prime }=\vartheta +\pi /2$ and $\varphi $ with the magnetic
colatitude and azimuth, respectively.

As shown in Section \ref{profile}, the pulse profiles based on the Stokes
parameters, written as functions of the magnetic azimuth $\varphi $ as given
by Eqs. (\ref{pp1})-(\ref{pp4}), can be plotted against the rotation phase $%
\phi $ instead of $\varphi $. This requires the transformation between $%
\varphi $\ and $\phi $, which is given through the analysis below.

In the corotating frame described in Appendix \ref{appA}, the magnetic
azimuth can be identified as $\varphi =\pi /2-\phi ^{\prime }$. Then by
means of Eqs. (\ref{x1}) and (\ref{y1}) together with Eqs. (\ref{tr2})-(\ref%
{tr4}), one can write down 
\begin{align}
\frac{x^{\prime }}{r}& =\sin \theta ^{\prime }\cos \phi ^{\prime }=\sin
\theta ^{\prime }\sin \varphi  \notag \\
& =\sin \theta \sin \left( \phi -\Omega t\right),  \label{ph1} \\
\frac{y^{\prime }}{r}& =\sin \theta ^{\prime }\sin \phi ^{\prime }=\sin
\theta ^{\prime }\cos \varphi  \notag \\
& =\sin \alpha \cos \theta -\cos \alpha \sin \theta \cos \left( \phi -\Omega
t\right) .  \label{ph2}
\end{align}%
Combining these two, we obtain the conversion expression between $\varphi $
and $\phi $: 
\begin{equation}
\varphi =\arctan \left( \frac{\sin \theta \sin \left( \phi -\Omega t\right) 
}{\sin \alpha \cos \theta -\cos \alpha \sin \theta \cos \left( \phi -\Omega
t\right) }\right) .  \label{ph3}
\end{equation}%
For radiation from a source charge at $\theta _{\mathrm{s}}=\alpha
+\varepsilon $ on a magnetic field line at $t=0$, the expression reduces to 
\begin{equation}
\varphi =\arctan \left( \frac{\sin \left( \alpha +\varepsilon \right) \sin
\phi }{\sin \alpha \cos \left( \alpha +\varepsilon \right) -\cos \alpha \sin
\left( \alpha +\varepsilon \right) \cos \phi }\right) ,  \label{ph4}
\end{equation}%
where $\alpha $ and $\varepsilon $ denote the inclination angle and the
sight line impact angle, respectively. It can be checked from (\ref{ph3})
that $\varphi \rightarrow -\phi $ in the alignment limit $\alpha \rightarrow
0$. Also, from (\ref{ph4}) it can be shown that for $\phi \ll 1$, 
\begin{equation}
\varphi \simeq -\frac{\sin \left( \alpha +\varepsilon \right) }{\sin
\varepsilon }\phi +\mathcal{O}\left( \phi ^{2}\right) .  \label{ph5}
\end{equation}

The pulse profiles (\ref{pp1})-(\ref{pp4}) are dependent upon the curvature
radius $\rho $ through $\omega _{\mathrm{o}}$ given by Eq. (\ref{sp17}).
Along with the conversion expression (\ref{ph4}) above, the curvature radius
as given by Eq. (\ref{sp14}) should also be reexpressed. By means of Eq. (%
\ref{tr5}), with $\theta _{\mathrm{s}}=\alpha +\varepsilon $, the magnetic
colatitude $\theta _{\mathrm{s}}^{\prime }$ is reduced to 
\begin{equation}
\cos \theta _{\mathrm{s}}^{\prime }=\cos \alpha \cos \left( \alpha
+\varepsilon \right) +\sin \alpha \sin \left( \alpha +\varepsilon \right)
\cos \left( \phi -\Omega t\right) .  \label{rho1}
\end{equation}%
Plugging this into Eq. (\ref{sp14}), the curvature radius can be finally
written as 
\begin{widetext}
\begin{equation}
\rho =r_{\mathrm{s}}\left[ 1+\,\frac{g^{2}\left( r_{\mathrm{s}}\right) \left[
\cos \alpha \cos \left( \alpha +\varepsilon \right) +\sin \alpha \sin \left(
\alpha +\varepsilon \right) \cos \phi \right] ^{2}}{1-\left[ \cos \alpha
\cos \left( \alpha +\varepsilon \right) +\sin \alpha \sin \left( \alpha
+\varepsilon \right) \cos \phi \right] ^{2}}\right] ^{1/2}.  \label{rho2}
\end{equation}
\end{widetext}

\end{document}